# Cross-stream migration of a surfactant-laden deformable droplet in a Poiseuille flow

Sayan Das, Shubhadeep Mandal and Suman Chakraborty†

Department of Mechanical Engineering, Indian Institute of Technology Kharagpur,
Kharagpur – 721302, India

The motion of a viscous deformable droplet suspended in an unbounded Poiseuille flow in the presence of bulk-insoluble surfactants is studied analytically. Assuming the convective transport of fluid and heat to be negligible, we perform a small-deformation perturbation analysis to obtain the droplet migration velocity. The droplet dynamics strongly depends on the distribution of surfactants along the droplet interface, which is governed by the relative strength of convective transport of surfactants as compared with the diffusive transport of surfactants. The present study is focused on the following two limits: (i) when the surfactant transport is dominated by surface diffusion, and (ii) when the surfactant transport is dominated by surface convection. In the first limiting case, it is seen that the axial velocity of the droplet decreases with increase in the advection of the surfactants along the surface. The variation of cross-stream migration velocity, on the other hand, is analyzed over three different regimes based on the ratio of the viscosity of the droplet phase to that of the carrier phase $(\lambda)$. In the first regime $(\sim \lambda < 0.7)$ the migration velocity decreases with increase in surface advection of the surfactants although there is no change in direction of droplet migration. For the second regime $(\sim 0.75 < \lambda < 11)$, the direction of the cross-stream migration of the droplet changes depending on different parameters. In the third regime $(\sim \lambda > 11)$, the migration velocity is merely affected by any change in the surfactant distribution. For the other limit of higher surface advection in comparison to surface diffusion of the surfactants, the axial velocity of the droplet is found to be independent of the surfactant distribution. However, the cross-stream velocity is found to decrease with increase in non-uniformity in surfactant distribution.

# I. INTRODUCTION

The dynamics of deformable droplets suspended in a pressure driven flow has been of great interest to the scientific community due to its wide applications in different microfluidic devices.[1–3] Some of the common usages of this droplet based devices includes cell encapsulation, reagent mixing, analytic detection and drug delivery.[1,4–7] Surface active agents are quite common in microfluidic devices, which are found useful for the purpose of droplet generation, stabilization of various emulsions as well as controlling the properties of polymer blends and emulsions.[8–11] The surface tension at the droplet interface is found to decrease in regions of high surfactant concentration on the droplet surface.[12] Thus a thorough understanding of the dynamics of a surfactant laden drop is required in order to optimize various flow characteristics.

The study of droplet dynamics in a pressure driven flow has been the focus of research, both theoretically and experimentally, since a long time.[13–15] Haber and Hetsroni first studied the droplet dynamics of a non-deformable Newtonian droplet with a clean interface suspended in another Newtonian fluid with an imposed Poiseuille flow.[16] However, the effect of different non-linear effects like deformation, inertia[14,17,18] or fluid viscoelasticity[19] on droplet dynamics were later on investigated. It has been shown that an eccentrically placed deformable droplet, in the absence of any other non-linear effects migrates axially as well as in the cross-stream direction.[16,19–23] Chan et al., in their work, have consider both the effects of deformability and viscoelasticity.[19] They have shown that a Newtonian deformable droplet, which when suspended in a second-order fluid undergoing Poiseuille flow, migrates away from the flow centerline for $0.5 < \lambda < 10$, where $\lambda$ is the ratio of the droplet phase viscosity to the viscosity of the suspending phase. For any other values of $\lambda$, the droplet always migrates towards the channel centerline.

For a surfactant-free or a clean droplet, the droplet deformation is solely controlled by the ratio of the viscous forces to the surface tension forces acting on the droplet.[24,25] However, in the presence of different contaminants or surfactants on the droplet interface, surface tension is reduced and may become non-uniform depending on distribution of the surfactants. Previously performed experiments show that there exists a relationship between shape deformation of droplet, surfactant redistribution on droplet surface and bulk flows.[10,26–30] Quite a number of theoretical studies have considered the dynamics of a surfactant laden spherical droplet. Haber and Hetsroni in their analysis derived the expression for the terminal settling velocity of a surfactant covered droplet suspended in an arbitrary flow.[31] They performed only a leading order analysis without consideration of any correction to the spherical shape of the droplet. Later Hanna and Vlahovska considered a surfactant-laden droplet in a Poiseuille flow.[32] They showed that the droplet migrates towards the flow centerline due to the flow-induced asymmetry in surfactant concentration along the surface of the droplet. They performed an asymptotic analysis for the limiting case in which the surfactant transport along the droplet surface is dominated by surface advection. Pak et al. recently in their work did a similar asymptotic analysis for the limiting case of surface diffusion dominated surfactant transport. In either of the studies no droplet deformation was taken into consideration.[33]

It has been shown experimentally as well as numerically by Stan et al. that deformation plays an important role in determining the dynamics of a droplet.[13] They showed that deformation-induced lift force affects the lateral migration characteristics of the droplet. Vlahovska et al. in their study used a small-deformation asymptotic method to analyze the effect of surfactant induced Marangoni stress on the dynamics of a droplet suspended in a linear flow.[34] They took into account a shape correction to the spherical droplet and performed the study for the limiting case when the main mode of surfactant transport along the droplet surface is by advection. However, any study on the dynamics of a surfactant-laden deformable droplet suspended in a Poiseuille flow is missing in the literature. In the present study, we see how the deformation as well as bulk flow induced surfactant redistribution on the droplet surface affects the axial as well as the lateral migration of the droplet. Towards this, we use a small-deformation asymptotic analysis to tackle the non-linearity present in the problem due to the coupled set of non-linear governing equations. The transport of surfactants along the droplet surface due to advection couples surfactant distribution with the flow field. Neglecting fluid inertia and thermal convection, we thus perform an asymptotic analysis for two different limits: (i) when the surfactant transport along the droplet surface is dominated by surface diffusion and (ii) when the surfactant transport is dominated by convection. The presence of surfactants on the droplet surface not only alters the magnitude of droplet velocity, but also changes the direction of migration, irrespective of $\lambda$.

## II. PROBLEM FORMULATION

### A. System description

The system under analysis consists of a neutrally buoyant Newtonian droplet of undeformed radius $a$ and having density $\rho$ and viscosity $\mu_i$. The droplet is suspended in another Newtonian medium of density and viscosity of $\rho$ and $\mu_e$ respectively. Surfactants, insoluble in either of the phases are assumed to be present at the interface of the droplet. The advection as well as diffusion of the surfactants takes place along the droplet surface. A schematic of the physical system is provided in Fig. 1. The change in concentration of the surfactant molecules on the droplet surface, due to the presence of the imposed Poiseuille flow, alters the interfacial tension $\left(\bar{\sigma}\right)$. This surface tension depends solely on the surfactant distribution $\left(\bar{\Gamma}\right)$ along the droplet interface. All the other material properties in the present analysis are assumed to be constant. The uniform surfactant distribution on the surface of the droplet, suspended in a quiescent medium, is denoted by $\bar{\Gamma}_{eq}$. The corresponding surface tension for the case of a surfactant-free droplet is denoted by $\bar{\sigma}_{eq}$. This equilibrium surfactant distribution is disturbed due the presence of the imposed Poiseuille flow $(\bar{\mathbf{V}}_\infty)$ which causes advection of the surfactant molecules. The resulting non-uniform distribution of surfactants along the interface generates Marangoni stress which causes deformation of the droplet as well as alters its migration velocity. The aim of the present study is to investigate the combined effect of an imposed Poiseuille flow, shape deformation and surfactant distribution on the migration characteristics of the droplet. We, in our analysis, have considered a spherical coordinate system $(\bar{r}, \theta, \varphi)$ attached to the centroid of the undeformed spherical droplet (see Fig.1).

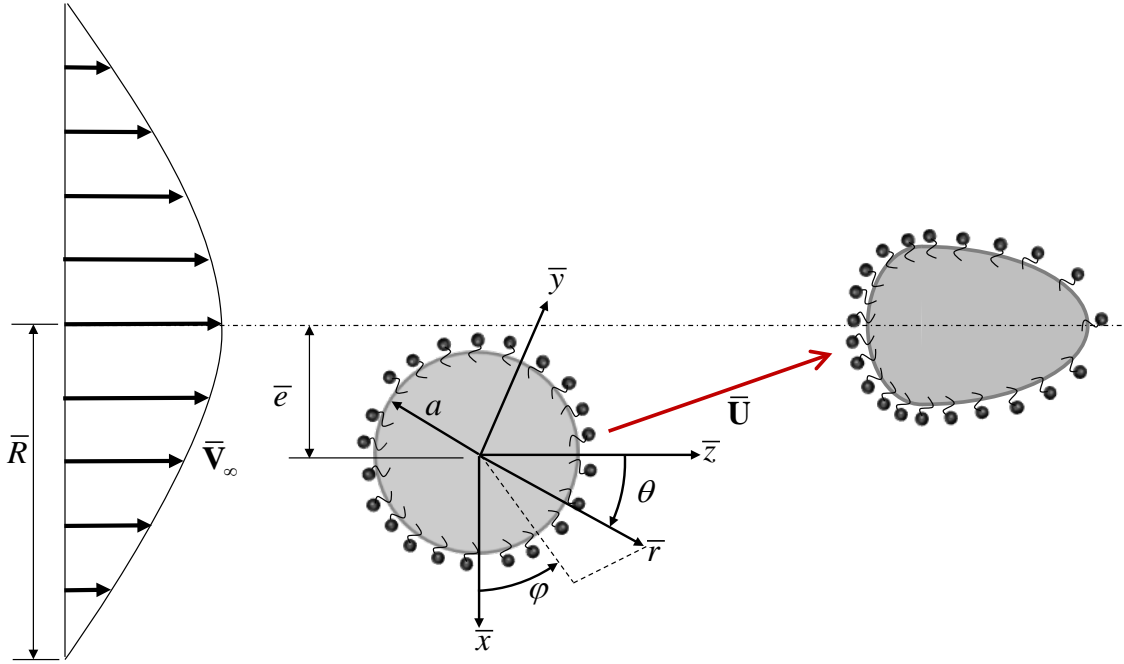

Fig. 1. Schematic of a droplet of radius $a$ suspended in an cylindrical Poiseuille flow $(\bar{\mathbf{V}}_\infty)$. The droplet is initially placed in an eccentric position at a distance $\bar{e}$ from the centerline of flow. Both the spherical $(\bar{r}, \theta, \varphi)$ and the Cartesian $(\bar{x}, \bar{y}, \bar{z})$ coordinates system are shown, with the $\bar{x}$ axis being directed away from the flow centerline.

## B. Assumptions

We make use of the following assumptions to simplify the governing equations and their corresponding boundary conditions.

(i) The inertial effect on the flow field is neglected. The flow field is assumed to be dominated by the viscous and the pressure forces, that is the hydrodynamic Reynolds number $\left(Re = \rho \bar{V}_c a / \mu_e\right)$ is taken to small enough. Here $\bar{V}_c$ is the centerline velocity of the imposed Poiseuille flow.

(ii) The surfactant is assumed to be bulk-insoluble.[35]

(iii) A linear dependence of the interfacial tension on the surfactant concentration is assumed.[36,37]

(iv) The droplet is assumed to be suspended in an unbounded domain. That is the effect of bounding walls, if present, are assumed to be negligible due to small size of the droplet.

(v) Only small deformation of the droplet is considered. This is possible when the surface tension force dominates the viscous forces trying to deform the droplet. This can be mathematically represented as $Ca^* \ll 1$, where $Ca^* = \mu_e \bar{V}_c / \bar{\sigma}_{eq}$ is the Capillary number denoting the ratio of the viscous forces to the surface tension forces.

## C. Governing equations and boundary conditions

The flow field is governed by the continuity and Navier-Stokes equations. However, due to the assumption of small Reynolds number, the inertia terms are neglected and the governing equation reduces to the simplified continuity and Stokes equations. The governing equations for both the phases, inside as well as outside the droplet are provided below

$$\left.\begin{array}{l} -\bar{\nabla} \bar{p}_i + \mu_i \bar{\nabla}^2 \bar{\mathbf{u}}_i = \mathbf{0}, \ \bar{\nabla} \cdot \bar{\mathbf{u}}_i = 0, \\ -\bar{\nabla} \bar{p}_e + \mu_e \bar{\nabla}^2 \bar{\mathbf{u}}_e = \mathbf{0}, \ \bar{\nabla} \cdot \bar{\mathbf{u}}_e = 0, \end{array}\right\} \quad (1)$$

with $(\bar{\mathbf{u}}, p)$ being the velocity and pressure fields respectively. The subscripts $(i, e)$ are used to denote quantities inside and outside the droplet respectively. The far-field condition satisfying the above set of governing equations are written below

$$\begin{array}{ll} \text{as } \bar{r} \to \infty, & \bar{\mathbf{u}}_e = \bar{\mathbf{V}}_\infty - \bar{\mathbf{U}}, \\ \text{as } \bar{r} \to \infty, & \bar{p}_e = \bar{p}_\infty, \end{array} \quad (2)$$

where $\bar{\mathbf{U}}$ is the droplet migration velocity vector, $\left(\bar{\mathbf{V}}_\infty, \bar{p}_\infty\right)$ are the velocity and pressure at the far-field. The far-field imposed velocity profile is written below in accordance to the spherical coordinate system (attached to the centroid of the droplet) in the following form

$$\bar{\mathbf{V}}_\infty = \bar{V}_c \left[1 - \frac{\bar{e}^2}{\bar{R}^2} - \frac{\bar{r}^2}{\bar{R}^2}\sin^2\theta - \frac{2\bar{r}}{\bar{R}^2}\bar{e}\cos\varphi\sin\theta\right]\mathbf{e}_z, \quad (3)$$

where, $\bar{R}$ denotes the position of the zero imposed velocity with respect the centerline of flow. The velocity as well as the pressure fields inside the droplet, $(\bar{\mathbf{u}}_i, \bar{p}_i)$ is bounded at the centroid of the droplet, $\bar{r} = 0$. Other than these, the boundary conditions at the droplet interface for the flow field are the kinematic boundary condition (or the no penetration boundary condition), the tangential velocity continuity condition and the condition for balance between the hydrodynamic and surfactant Marangoni stress. These boundary conditions are written below, respectively, in the following form

$$\left. \begin{array}{l} \text{at } \bar{r} = \bar{r}_s, \quad \bar{\mathbf{u}}_i = \bar{\mathbf{u}}_e, \\ \text{at } \bar{r} = \bar{r}_s, \quad \bar{\mathbf{u}}_i \cdot \mathbf{n} = \bar{\mathbf{u}}_e \cdot \mathbf{n} = 0, \\ \text{at } \bar{r} = \bar{r}_s, \quad (\bar{\tau}_e \cdot \mathbf{n} - \bar{\tau}_i \cdot \mathbf{n}) = -\bar{\nabla}_s \bar{\sigma} + \bar{\sigma}(\bar{\nabla} \cdot \mathbf{n})\mathbf{n}, \end{array} \right\} \quad (4)$$

where, $\bar{r}_s$ represents the surface of the droplet and $\bar{\nabla}_s = (\mathbf{I} - \mathbf{nn}) \cdot \bar{\nabla}$ is the surface gradient operator. The total stress tensors inside as well as outside the droplet are given by $\bar{\tau}_i = -\bar{p}_i \mathbf{I} + \mu_i \left[ \bar{\nabla} \bar{\mathbf{u}}_i + (\bar{\nabla} \bar{\mathbf{u}}_i)^T \right]$ and $\bar{\tau}_e = -\bar{p}_e \mathbf{I} + \mu_e \left[ \bar{\nabla} \bar{\mathbf{u}}_e + (\bar{\nabla} \bar{\mathbf{u}}_e)^T \right]$. $\mathbf{n}$ is the unit normal to the surface of the droplet and is given by

$$\mathbf{n} = \frac{\bar{\nabla} F}{|\bar{\nabla} F|}, \quad (5)$$

where, $F = \bar{r} - \bar{r}_s$ is the equation of the surface of the droplet.

The Marangoni stress is dependent on the surface tension of the droplet $(\bar{\sigma})$. As per our assumption, the surface tension depends linearly on the surfactant concentration and can be represented through an equation of state in the following form[36,37]

$$\bar{\sigma} = \bar{\sigma}_{eq} - R_g \bar{T}_o \bar{\Gamma}, \quad (6)$$

where $\bar{T}_o$ is any reference temperature and $R_g$ is the universal gas constant.

The local surfactant concentration $\bar{\Gamma}$ is governed by the surfactant transport equation and can be written as[26]

$$\bar{\nabla}_s \cdot (\bar{\mathbf{u}}_s \bar{\Gamma}) = D_s \bar{\nabla}_s^2 \bar{\Gamma}, \quad (7)$$

where, $D_s$ is the surface diffusivity of the surfactants and $\bar{\mathbf{u}}_s$ is the interfacial fluid velocity.

With the dimensional governing equations and boundary conditions at hand we move forward to expressing these in their non-dimensional form. Towards this we first define a non-dimensional scheme as follows

$$\left. \begin{array}{l} r = \bar{r}/a,\ \mathbf{u} = \bar{\mathbf{u}}/\bar{V}_c,\ \Gamma = \bar{\Gamma}/\bar{\Gamma}_{eq},\ \sigma = \bar{\sigma}/\bar{\sigma}_{eq}, \\ p = \bar{p}/(\mu_e \bar{V}_c/a),\ \boldsymbol{\tau} = \bar{\boldsymbol{\tau}}/(\mu_e \bar{V}_c/a) \end{array} \right\} \quad (8)$$

It should be kept in mind that all dimensional quantities are denoted by an 'overbar', whereas for the non-dimensional quantities as well as the material properties no 'overbar' is used. The other non-dimensional entities encountered while deriving the dimensionless set of governing equations and boundary conditions are: (i) the viscosity ratio, $\lambda = \mu_i/\mu_e$, (ii) the elasticity parameter, $\beta = \bar{\Gamma}_{eq} R T_o / \bar{\sigma}_{eq}$, which indicates the sensitivity of the surface tension to a change in the local surfactant concentration, (iii) the modified capillary number, $Ca = Ca^*/(1-\beta)$, and (iv) the surface Péclet number, $Pe_s = \bar{V}_c a/D_s$, which signifies the relative strength of the transport of surfactants due to convection as compared to that due to surface diffusion. From equation (6) it is clearly seen that $\beta = -d(\bar{\sigma}/\bar{\sigma}_{eq})/d\bar{\Gamma}$, thus inferring that the equilibrium surface tension for a surfactant-laden droplet in the absence of any bulk flow is given by $\bar{\sigma}_{eq}(1-\beta)$ corresponding to a uniform surfactant concentration of $\bar{\Gamma}_{eq}$. It is thus more convenient to consider the capillary number based on the equilibrium surface tension of a surfactant laden drop rather than that of a clean drop. This is the reason for the usage of a modified capillary number.

With the help of the above mentioned non-dimensional scheme, the dimensionless governing equations for the velocity and the pressure field can be rewritten in the following form

$$\left. \begin{array}{l} -\nabla p_i + \lambda \nabla^2 \mathbf{u}_i = \mathbf{0},\ \nabla \cdot \mathbf{u}_i = 0, \\ -\nabla p_e + \nabla^2 \mathbf{u}_e = \mathbf{0},\ \nabla \cdot \mathbf{u}_e = 0, \end{array} \right\} \quad (9)$$

and the corresponding non-dimensional boundary conditions are given by

$$\left. \begin{array}{l} \text{at } r \to \infty,\ (\mathbf{u}_e, p_e) = (\mathbf{V}_\infty - \mathbf{U}, p_\infty), \\ \mathbf{u}_i \text{ is bounded at } r = 0, \\ \text{at } r = r_s,\ \mathbf{u}_i \cdot \mathbf{n} = \mathbf{u}_e \cdot \mathbf{n} = 0, \\ \text{at } r = r_s,\ \mathbf{u}_i = \mathbf{u}_e, \\ \text{at } r = r_s,\ (\boldsymbol{\tau}_e \cdot \mathbf{n} - \boldsymbol{\tau}_i \cdot \mathbf{n}) = \dfrac{\beta}{(1-\beta)Ca} \nabla_s \Gamma + \dfrac{\sigma}{Ca}(\nabla \cdot \mathbf{n})\mathbf{n}. \end{array} \right\} \quad (10)$$

The last boundary condition which is the stress balance condition is obtained by usage of the non-dimensional form of the equation of state given by

$$\sigma = 1 - \beta \Gamma \quad (11)$$

It should be kept in mind that the surface tension in the above equation is based on the modified capillary number and can be expressed as

$$\sigma = \frac{\bar{\sigma}}{\bar{\sigma}_{eq}(1-\beta)}. \quad (12)$$

This shows that $\beta$ is bounded between 0 and 1. Finally, the surfactant transport equation in its dimensionless form is represented as

$$Pe_s \nabla_s \cdot (\mathbf{u}_s \Gamma) = \nabla_s^2 \Gamma. \quad (13)$$

The mass conservation constraint for the surfactants present on the droplet surface has to be also fulfilled while solving for the surfactant concentration and is provided below

$$\int_{\varphi=0}^{2\pi} \int_{\theta=0}^{\pi} \Gamma(\theta, \varphi) \sin\theta \, d\theta \, d\varphi = 4\pi. \quad (14)$$

Looking into equation (13), it is seen from the non-linear surfactant convection term on the left hand side of the same equation that the flow field and surfactant transport are coupled and has to be solved simultaneously. The solution of the flow field as well as the surfactant concentration is analytically not possible from the above set of equations in its present form. Thus an asymptotic approach is used to solve for the flow field.[32,33] This asymptotic analysis is done for the following two special limiting cases: (i) Low surface Péclet number, $Pe_s \ll 1$, which signifies that the dominant mode of surfactant transport along the droplet surface is by surface diffusion. It physically signifies a large value of surface diffusivity of the surfactants. (ii) Large surface Péclet number, $Pe_s \gg 1$, denoting that convection is the main mode of surfactant transport. This physically indicates a low value of surface diffusivity.[34]

### III. ASYMPTOTIC SOLUTION

We first represent the different field variables in terms of spherical solid harmonics. We start with the velocity and pressure fields in either of the phases. The velocity and the pressure fields inside the droplet satisfy the Stokes equation and thus the general Lamb's solution can be used to express them in terms of growing spherical solid harmonics as follows

$$\left.\begin{aligned}\mathbf{u}_i &= \sum_{n=1}^{\infty}\left[\nabla\times(\mathbf{r}\chi_n)+\nabla\Phi_n+\frac{n+3}{2(n+1)(2n+3)\lambda}r^2\nabla p_n-\frac{n}{(n+1)(2n+3)\lambda}\mathbf{r}p_n\right],\\ p_i &= \sum_{n=0}^{\infty}p_n,\end{aligned}\right\} \quad (15)$$

where, $p_n$, $\Phi_n$ and $\chi_n$ are growing spherical solid harmonics which can be expressed as

$$\left.\begin{aligned}p_n &= \lambda r^n\sum_{m=0}^{n}\left[A_{n,m}\cos(m\varphi)+\hat{A}_{n,m}\sin(m\varphi)\right]P_{n,m}(\cos\theta),\\ \Phi_n &= r^n\sum_{m=0}^{n}\left[B_{n,m}\cos(m\varphi)+\hat{B}_{n,m}\sin(m\varphi)\right]P_{n,m}(\cos\theta),\\ \chi_n &= r^n\sum_{m=0}^{n}\left[C_{n,m}\cos(m\varphi)+\hat{C}_{n,m}\sin(m\varphi)\right]P_{n,m}(\cos\theta).\end{aligned}\right\} \quad (16)$$

where, $P_{n,m}(\cos\theta)$ are associated Legendre polynomials of degree $n$ and order $m$. In a similar manner, the velocity and the pressure fields outside the drop can be expressed as combination of the decaying solid spherical harmonics and the far field quantities. A proper representation of the outer phase flow variables is given below

$$\left.\begin{aligned}\mathbf{u}_e &= (\mathbf{V}_\infty-\mathbf{U})+\sum_{n=1}^{\infty}\left[\nabla\times(\mathbf{r}\chi_{-n-1})+\nabla\Phi_{-n-1}-\frac{n-2}{2n(2n-1)}r^2\nabla p_{-n-1}+\frac{n+1}{n(2n-1)}\mathbf{r}p_{-n-1}\right],\\ p_e &= p_\infty+\sum_{n=0}^{\infty}p_{-n-1},\end{aligned}\right\} \quad (17)$$

where $p_{-n-1}$, $\Phi_{-n-1}$ and $\chi_{-n-1}$ are the decaying solid harmonics and are given by

$$\left.\begin{aligned}p_{-n-1} &= r^{-n-1}\sum_{m=0}^{n}\left[A_{-n-1,m}\cos(m\varphi)+\hat{A}_{-n-1,m}\sin(m\varphi)\right]P_{n,m}(\cos\theta),\\ \Phi_{-n-1} &= r^{-n-1}\sum_{m=0}^{n}\left[B_{-n-1,m}\cos(m\varphi)+\hat{B}_{-n-1,m}\sin(m\varphi)\right]P_{n,m}(\cos\theta),\\ \chi_{-n-1} &= r^{-n-1}\sum_{m=0}^{n}\left[C_{-n-1,m}\cos(m\varphi)+\hat{C}_{-n-1,m}\sin(m\varphi)\right]P_{n,m}(\cos\theta).\end{aligned}\right\} \quad (18)$$

Towards calculating the velocity and the pressure fields, the constant coefficients, $A_{n,m}$, $B_{n,m}$, $C_{n,m}$, $A_{-n-1,m}$, $B_{-n-1,m}$, $C_{-n-1,m}$, $\hat{A}_{n,m}$, $\hat{B}_{n,m}$, $\hat{C}_{n,m}$, $\hat{A}_{-n-1,m}$, $\hat{B}_{-n-1,m}$ and $\hat{C}_{-n-1,m}$, are found out from the remaining boundary conditions at the droplet interface namely the kinematic condition, the tangential velocity continuity condition and the tangential stress balance condition. The tangential component of stress balance can be obtained from the stress balance condition given in equation (10). This condition is due to variation of surface tension caused by the non-uniform distribution of surfactants along the droplet surface which generates a Marangoni stress. This causes a stress discontinuity or jump at the interface. The tangential stress boundary condition can thus be expressed as

$$\text{at } r = r_s, \ (\boldsymbol{\tau}_e \cdot \mathbf{n} - \boldsymbol{\tau}_i \cdot \mathbf{n}) \cdot (\mathbf{I} - \mathbf{nn}) = \frac{\beta}{(1-\beta)Ca} (\nabla_s \Gamma) \cdot (\mathbf{I} - \mathbf{nn}). \tag{19}$$

Finally, the surfactant concentration can be expressed in terms of solid spherical harmonics as

$$\Gamma = \sum_{n=0}^{\infty} \sum_{m=0}^{n} \left[ \Gamma_{n,m} \cos(m\varphi) + \hat{\Gamma}_{n,m} \sin(m\varphi) \right] P_{n,m}(\cos\theta), \tag{20}$$

Here the constant coefficients $\Gamma_{n,m}$ and $\hat{\Gamma}_{n,m}$ are found out by solving the surfactant transport equation along with the governing equations for the flow field.

## A. Solution for $Pe_s \ll 1$

In this limit we assume the surface Péclet number to be of the same order as that of the Capillary number, $Ca$. That is $Pe_s \sim Ca$. In a more mathematical format, this can be written in the following form

$$Pe_s = kCa, \tag{21}$$

where $k = a\bar{\sigma}_{eq}(1-\beta)/\mu_e D_s$ is a quantity of finite magnitude $(\sim O(1))$ and depends on the material properties solely. Thus for a given value of $k$ and $\beta$, droplet deformation is a function of the Capillary number only. Hence capillary number is chosen as the perturbation parameter for the asymptotic analysis. In this limit, we expand all flow variables in power series of $Ca$ in the following form

$$\psi = \psi^{(0)} + \psi^{(Ca)}Ca + O(Ca^2), \tag{22}$$

where, $\psi$ denotes any generic flow variable. The first term on the right hand side of the above equation denotes the leading order term that considers zero deformation of the droplet. The other terms indicate the correction terms of $O(Ca)$ or $O(Ca^2)$ due to deformation of the droplet. The local surfactant concentration, $\Gamma$, on the other hand, is expanded in the following form[34]

$$\Gamma = 1 + \Gamma^{(0)}Ca + \Gamma^{(Ca)}Ca^2 + O(Ca^3). \tag{23}$$

In order to satisfy equation (14) or the mass conservation constraint for the surfactants, the first term in the expansion for surfactant concentration is taken as 1. The leading order surfactant transport equation can be obtained by substituting equations (22) and (23) into equation (13) and is written as

$$\nabla_s^2 \Gamma^{(0)} = k \nabla_s \cdot \mathbf{u}_s^{(0)}. \tag{24}$$

We follow a strategic methodology to obtain the droplet migration velocity at different orders of perturbation.

(i) We first substitute equations (21), (22) and (23) into the leading order governing equations and boundary conditions for flow field, that is equations (9), (10) and (13) to obtain the leading order governing equations and boundary conditions.

(ii) Next the leading order flow field boundary conditions (except the normal stress balance) and the surfactant transport equation (equation (24)) are solved simultaneously to obtain the constant coefficients present in equation (16) and (18) as well as the local surfactant concentration for the leading order. From the expressions of these constant coefficients, the solid spherical harmonics are found out and substituted in equation (15) and (17) to obtain the leading order velocity and pressure fields. The expression of the leading order surfactant concentration, subsequently found out is given below

$$\Gamma^{(0)} = \Gamma_{1,0}^{(0)} P_{1,0} + \Gamma_{2,1}^{(0)} \cos\varphi P_{2,1} + \Gamma_{3,0}^{(0)} P_{3,0}, \tag{25}$$

where the expressions of the constant coefficients $\Gamma_{1,0}^{(0)}$, $\Gamma_{3,0}^{(0)}$ and $\Gamma_{2,1}^{(0)}$ are provided in Appendix A.

(iii) The droplet migration velocity, both lateral and axial, at this order is found out by using the force-free condition. This condition is obtained by letting the net drag force acting on the droplet equal to zero and can be written as

$$\mathbf{F}_H^{(0)} = 4\pi \nabla \left( r^3 p_{-2}^{(0)} \right) = \mathbf{0}, \tag{26}$$

where, $p_{-2}^{(0)}$ is a leading order decaying solid harmonic of degree 1. Thus the expression of the leading order solution for the droplet migration velocity is found out and is stated below

$$U_z^{(0)} = \left[\left\{1 - \frac{e^2}{R^2} - \frac{2\lambda}{(2+3\lambda)R^2}\right\} - \underbrace{\frac{4}{3(2+3\lambda)R^2}\left\{\frac{\beta k}{(k-3\lambda-2)\beta + 3\lambda + 2}\right\}}_{\text{correction due to presence of surfactants}}\right], \qquad (27)$$

$$U_x^{(0)} = U_y^{(0)} = 0.$$

From the above expression, it is seen that the second term in the expression of the axial migration velocity above represents the correction term due to the presence of a non-uniform distribution of surfactants. It can be clearly seen that by substituting $\beta = 0$ in the above expression, we obtain the migration velocity for a clean droplet.

(iv) Upon obtaining the leading order solution, the normal stress balance at the deformed interface, $r = r_s = 1 + Ca g^{(Ca)}(\theta,\varphi) + Ca^2 g^{(Ca^2)}(\theta,\varphi)$, is used to calculate the $O(Ca)$ correction to the spherical shape of the droplet. Here $g^{(Ca)}$ and $g^{(Ca^2)}$ are the $O(Ca)$ and $O(Ca^2)$ correction to the droplet shape. We make use of the orthogonality condition for associate Legendre polynomials on either sides of the normal stress balance to find $g^{(Ca)}$. The expression for $g^{(Ca)}$, thus calculated, is given below

$$g^{(Ca)} = L_{2,1}^{(Ca)} \cos\varphi P_{2,1} + L_{3,0}^{(Ca)} P_{3,0}, \qquad (28)$$

where, the constant coefficients $L_{2,1}^{(Ca)}$ and $L_{3,0}^{(Ca)}$ are provided in Appendix A.

(v) Now that we have found out the $O(Ca)$ deformation of the droplet, we proceed further and derive all the flow field boundary conditions and the surfactant transport equation at the deformed interface of the droplet, $r = r_s$. The $O(Ca)$ surfactant transport equation thus found out is given below

$$\nabla_s^2 \Gamma^{(Ca)} = k \nabla_s \cdot \left\{ \mathbf{u}_s^{(Ca)} + \Gamma^{(0)} \mathbf{u}_s^{(0)} \right\}. \qquad (29)$$

(vi) The above equation along with the $O(Ca)$ flow boundary conditions are again solved simultaneously to obtain the velocity and pressure fields for both the phases as well as the surfactant concentration for $O(Ca)$ at the deformed surface. The expression for the $O(Ca)$ local surfactant concentration can be written as

$$\Gamma^{(Ca)} = \begin{Bmatrix} \Gamma^{(Ca)}_{0,0} + \Gamma^{(Ca)}_{1,1} \cos\varphi P_{1,1} + \Gamma^{(Ca)}_{2,0} P_{2,0} + \Gamma^{(Ca)}_{2,2} \cos\varphi P_{2,2} \\ +\Gamma^{(Ca)}_{3,1} \cos\varphi P_{3,1} + \Gamma^{(Ca)}_{4,0} P_{4,0} + \Gamma^{(Ca)}_{4,2} \cos 2\varphi P_{4,2} \end{Bmatrix}, \quad (30)$$

where, the constant coefficients, $\Gamma^{(Ca)}_{1,1}$, $\Gamma^{(Ca)}_{2,0}$, $\Gamma^{(Ca)}_{2,2}$, $\Gamma^{(Ca)}_{3,1}$, $\Gamma^{(Ca)}_{4,0}$ and $\Gamma^{(Ca)}_{4,2}$, in the above expression can be expressed in a general form as follows

$$\Gamma^{(Ca)}_{i,j} = \frac{a^{(Ca)}_{i,j}k^7 + b^{(Ca)}_{i,j}k^6 + c^{(Ca)}_{i,j}k^5 + d^{(Ca)}_{i,j}k^4 + f^{(Ca)}_{i,j}k^3 + h^{(Ca)}_{i,j}k^2 + l^{(Ca)}_{i,j}k + m^{(Ca)}_{i,j}}{q^{(Ca)}_{i,j}}. \quad (31)$$

The expressions of these constants are provided in Appendix A. The coefficient $\Gamma^{(Ca)}_{0,0}$ is found out from the surfactant conservation relation applied on the droplet interface as given in equation (14).

The $O(Ca)$ axial as well as the lateral components of droplet migration velocity are obtained from the force-free condition for this order, which can be written as

$$\mathbf{F}^{(Ca)}_H = 4\pi\nabla\left(r^3 p^{(Ca)}_{-2}\right) = \mathbf{0}, \quad (32)$$

where, $p^{(Ca)}_{-2}$ is an $O(Ca)$ decaying solid harmonic of degree 1. The expressions of the different components of $O(Ca)$ droplet migration velocity is given below

$$\left.\begin{matrix} U^{(Ca)}_z = U^{(Ca)}_y = 0, \\ U^{(Ca)}_x = \dfrac{c_1\beta^4 + c_2\beta^3 + c_3\beta^2 + c_4\beta + c_5}{c_6} e, \end{matrix}\right\} \quad (33)$$

where, the expression of the constant $c_1 - c_6$ are given in Appendix A. For a clean droplet $(\beta = 0)$ we get from the above equation $U^{(Ca)}_x = c_5/c_6$, which is found to match exactly with that obtained by Chan and Leal for Newtonian droplet.[19]

## B. Solution for $Pe_s \gg 1$

Contrary to the previous case for low surface Péclet number, we now assume that $Pe_s^{-1} \sim Ca$. We expand all the flow variables as well as the local surfactant concentration in a power series about $Ca$ similar to the case of $Pe_s \ll 1$ (see equations (22) and (23)). The only difference in this limiting case, as compared to the previous case is in the surfactant transport equation at different orders of perturbation.

We again highlight the key steps towards obtaining the droplet migration velocity in this limit.

(i) The leading order governing equations as well as boundary conditions for flow field are obtained in a manner similar as done in the limiting case of $Pe_s \ll 1$. The surfactant transport equation for leading order is given by

$$\nabla_s \cdot \mathbf{u}_s^{(0)} = 0. \tag{34}$$

(ii) The above equation along with the leading order flow field boundary conditions are solved simultaneously to obtain the solid spherical harmonics as well as the surfactant concentration. The surfactant concentration thus obtained for this order is given below

$$\left.\begin{aligned}\Gamma^{(0)} &= \Gamma_{1,0}^{(0)} P_{1,0} + \Gamma_{2,1}^{(0)} \cos\varphi P_{2,1} + \Gamma_{3,0}^{(0)} P_{3,0}, \\ \text{where,} \\ \Gamma_{1,0}^{(0)} &= -\frac{2}{R^2}\frac{1-\beta}{\beta}, \quad \Gamma_{2,1}^{(0)} = -\frac{5e}{3R^2}\frac{1-\beta}{\beta}, \quad \Gamma_{3,0}^{(0)} = \frac{7}{6R^2}\frac{1-\beta}{\beta}.\end{aligned}\right\} \tag{35}$$

(iii) Now with all the leading order solid harmonics known, we calculate the axial as well as the cross-stream component of the leading order droplet migration velocity. We again make use of the force-free condition as given in equation (26). The final expression of the droplet migration velocity is given below

$$\begin{aligned}U_z^{(0)} &= \left(1 - \frac{2}{3R^2}\right) - \frac{e^2}{R^2}, \\ U_x^{(0)} &= U_y^{(0)} = 0,\end{aligned} \tag{36}$$

It is clearly seen from the above expression that there is no contribution of surfactant distribution on the leading order migration velocity of the droplet.

(iv) The $O(Ca)$ deformation $\left(g^{(Ca)}\right)$ are next obtained by using the orthogonality conditions for associate Legendre polynomial in the normal stress balance condition on the deformed droplet surface. The expression for $O(Ca)$ correction to the droplet shape, $g^{(Ca)}$, is provided below

$$g^{(Ca)} = L_{2,1} \cos\varphi P_{2,1} + L_{3,0}^{(Ca)} P_{3,0},$$
where,
$$L_{2,1}^{(Ca)} = -\frac{5}{3}\frac{e}{R^2}, L_{3,0}^{(Ca)} = \frac{7}{12R^2}. \quad (37)$$

The above expressions for $O(Ca)$ deformation can also be derived from equation (28) in the limit $k \to \infty$.

(v) The $O(Ca)$ solution for flow field and surfactant concentration are obtained in a similar manner by simultaneously solving the flow boundary conditions and the surfactant transport equation, derived at the deformed interface of the droplet. The surfactant transport equation derived on the $O(Ca)$ deformed surface of the droplet, is given by

$$\nabla_s \cdot \left\{ \mathbf{u}_s^{(Ca)} + \Gamma^{(0)} \mathbf{u}_s^{(0)} \right\} = 0. \quad (38)$$

The $O(Ca)$ surfactant concentration thus obtained is written below in the following form

$$\Gamma^{(Ca)} = \begin{Bmatrix} \Gamma_{0,0}^{(Ca)} + \Gamma_{1,1}^{(Ca)} \cos\varphi P_{1,1} + \Gamma_{2,0}^{(Ca)} P_{2,0} + \Gamma_{2,2}^{(Ca)} \cos\varphi P_{2,2} \\ + \Gamma_{3,1}^{(Ca)} \cos\varphi P_{3,1} + \Gamma_{4,0}^{(Ca)} P_{4,0} + \Gamma_{4,2}^{(Ca)} \cos 2\varphi P_{4,2} \end{Bmatrix}, \quad (39)$$

where the expressions of the constant coefficients are provided in Appendix B. The general expression for $\Gamma_{0,0}^{(Ca)}$ as given in Appendix A remains the same.

(vi) The expressions for $O(Ca)$ droplet migration velocity can be obtained from the $O(Ca)$ force-free condition as was stated in equation (32) and is given below

$$U_z^{(Ca)} = 0, \ U_x^{(Ca)} = -\frac{e}{6R^4}\left(\frac{4}{\beta} - 3\right), \ U_y^{(Ca)} = 0. \quad (40)$$

As seen from the above expression, it can be inferred that the cross-stream migration velocity is always negative, that is, the droplet under the limit of high $Pe_s$ always migrates towards the flow centerline irrespective of the surfactant distribution along the surface of the droplet.

## IV. RESULTS AND DISCUSSION

The main result of our present analysis are the droplet migration velocities as well as the cross-stream trajectories of the surfactant laden droplet for two different asymptotic limits: (i) low surface Péclet limit and (ii) high surface Péclet limit. We first start our discussion on the asymptotic limit of low surface Péclet number.

## A. Low surface Péclet number Limit

The migration velocity of a surfactant covered droplet suspended in a Poiseuille flow is given by

$$\mathbf{U} = \left[ \left\{ 1 - \frac{e^2}{R^2} - \frac{2\lambda}{(2+3\lambda)R^2} \right\} - \underbrace{\frac{4}{3(2+3\lambda)R^2} \left\{ \frac{\beta k}{(k-3\lambda-2)\beta + 3\lambda + 2} \right\}}_{\text{correction due to presence of surfactants}} \right] \mathbf{e}_z \qquad (41)$$
$$+ Ca \left[ \frac{c_1\beta^4 + c_2\beta^3 + c_3\beta^2 + c_4\beta + c_5}{c_6} e \right] \mathbf{e}_x + O(Ca^2)$$

The above expression explicitly shows the contribution of non-uniform distribution of surfactants along the surface of the droplet. By substituting $\beta = 0$ in equation (41) we get the droplet migration velocity for a clean droplet. From equation (41) it can be said that surfactant induced Marangoni stress alters both the axial as well as the cross-stream migration velocity. It can be seen from equation (41) that the axial component of droplet migration velocity is independent of the droplet deformation when shape corrections till $O(Ca)$ is considered.

### 1. Effect of the property parameter, $k$

We first show the variation of the axial droplet migration velocity with viscosity ratio in Fig. 2. In each of the figures, 2(a) and 2(b), the variation of droplet migration velocity is shown for different values of the elasticity parameter, $(\beta = 0, 0.25, 0.5)$. In Fig. 2(a) the variation of the axial migration velocity of the droplet is shown for $k = 1$, while in Fig. 2(b) the variation is shown for $k = 3$. The other parameters involved in plot are mentioned in the caption of Fig. 2.

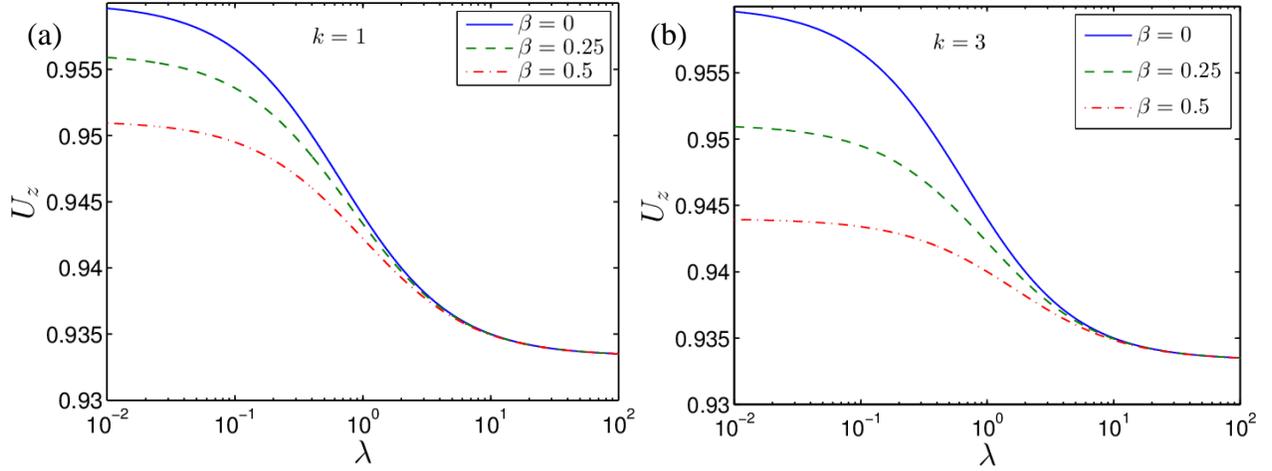

Fig. 2. Variation of axial migration velocity of the droplet with $\lambda$, for different values of $\beta$, that is $\beta = 0, 0.25, 0.5$. In Fig. (a) $k = 1$ and in Fig. (b) $k = 3$. The other parameters involved are $R = 5$, $e = 1$ and $Ca = 0.1$.

It is seen that the axial droplet migration velocity decreases as the viscosity ratio is increased. At higher viscosity ratios the droplet effectively behaves as a particle, that is, there is hardly any effect of the variation of surface tension on the droplet migration velocity and hence there is no effect of variation of $\beta$ or $k$. From the definition of $k$ we can write

$$k = \frac{Pe_s}{Ca} = \frac{a\bar{\sigma}_{eq}(1-\beta)}{\mu_e D_s}. \tag{42}$$

We can see from the above equation that both the property parameter, $k$, as well as the elasticity parameter, $\beta$, are coupled. Hence we show the effect of each of the parameters separately by varying one of the parameters while keeping the other parameter constant. With an effort to show the effect of the property parameter, at first, we keep the elasticity parameter constant at $\beta = 0.5$. Comparing Fig. 2(a) and Fig. 2(b), we see that increase in the property parameter from $k = 1$ to $k = 3$, reduces the axial migration velocity. The effect of $k$ is more prominent for the case of a bubble $(\lambda \to 0)$ as compared to the case of a droplet with a high value of $\lambda$.

A proper explanation about the above nature of variation of axial droplet velocity with $k$ can be given if we look into the variation of surfactant concentration and surface velocity about the droplet interface for different values of $k$ $(k = 1, 3)$ as shown in Fig. 3. Due to increase in $k$, the convective transport of surfactants along the interface of the droplet increases. As the fluid flow is from the front end of the droplet $(\theta = 0)$ to the rear end $(\theta = \pi)$ (refer to Fig. 3(b)), there is a higher concentration of surfactants at the rear end of the droplet compared to the front end and hence a lower surface tension at the rear end. Figure 3(a) confirms the fact that a higher value of $k$ results in a higher surfactant concentration at the rear end and a lower concentration at the front end. Thus a higher surface tension gradient due to a larger $k$ result in a larger Marangoni stress acting opposite to the direction of imposed flow. Hence the axial migration velocity reduces. This causes the surface velocity of the droplet to decrease as well.

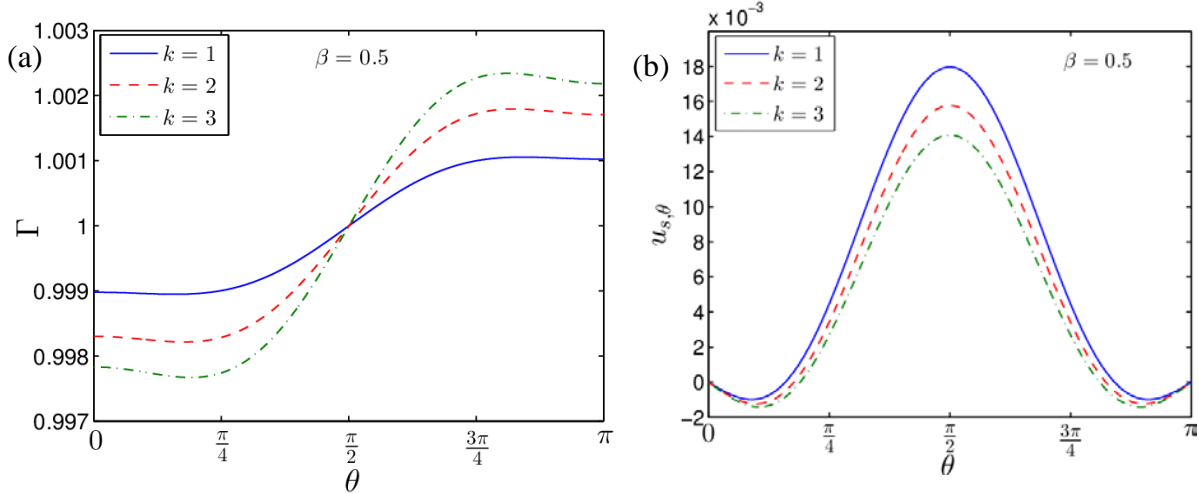

Fig. 3. (a) Variation of surfactant concentration with the polar angle $(\theta)$ about a particular transverse plane $(\varphi = \pi/2)$ for different values of $k(=1,2,3)$. (b) Variation of surface velocity with $\theta$ along a transverse plane, $\varphi = \pi/2$, and for $k = 1, 2, 3$. $u_{s,\theta}$ is the surface velocity at the droplet interface. The other parameters used in this plot are $R = 5$, $e = 1$, $\lambda = 1$, $\beta = 0.5$ and $Ca = 0.1$.

The variation of cross-stream migration velocity with the viscosity shows us some interesting outcomes. As above, we keep the value of $\beta$ fixed $(\beta = 0.5)$ and vary the property parameter $k$ $(k = 1, 3)$. As seen from Fig. 4(a) and 4(b), the cross-stream velocity of a low viscous droplet decreases with increase in the viscosity ratio whereas for a highly viscous droplet there is an increase in the cross-stream velocity. For $k = 1$, a droplet with low viscosity is seen to migrate towards the centerline of flow. As $\lambda$ increases there reaches a point $(\lambda = 0.75)$ where the direction of droplet migration changes. It starts migrating away from the flow centerline. On further increase in $\lambda$, it again reaches a point where the droplet changes its trajectory again and starts moving towards the flow centerline. The change in the direction of motion of the droplet occurs at a much higher value of viscosity ratio $(\lambda = 20)$. Now on comparison of figures 4(a) and 4(b), it is seen that an increase in the value of the property parameter, $k$ always results in a reduction in the magnitude of the cross-stream migration velocity of the droplet, provided $\beta$ is kept constant. If we consider the case for $\beta = 0.5$, then looking into Fig. 4(a), we see that for $k = 1$ a droplet with $\lambda = 1$ moves away from the flow centerline. However, for $k = 3$ (Fig. 4(b)), the same droplet $(\lambda = 1)$ continues to move towards the flow centerline. Thus variation in $k$ can cause change in the direction of motion of the droplet. The effect of variation of $k$ on the lateral migration velocity of the droplet is seen to be predominant for low viscous droplets or bubbles.

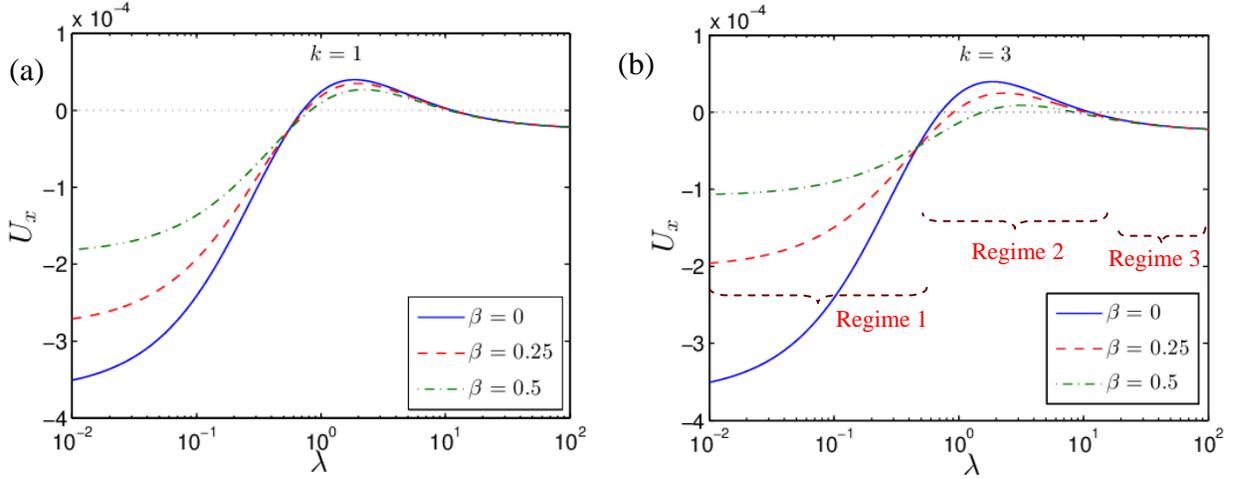

Fig. 4. Variation of cross-stream migration velocity of the droplet with $\lambda$, for different values of $\beta$, that is $\beta = 0, 0.25, 0.5$. In Fig. (a) $k = 1$ and in Fig. (b) $k = 3$. The other parameters involved are $R = 5$, $e = 1$ and $Ca = 0.1$.

The cross-stream migration owes its origin to the asymmetry in surfactant distribution about the axial plane. As the droplet is placed in an eccentric location with respect to the flow centerline, the portion of the droplet in the vicinity of the centerline has a higher surface velocity as compared to the portion away from the centerline. If the droplet is placed below the centerline (refer to Fig. 1), the upper hemisphere of the droplet has a higher surface velocity as compared to lower hemisphere. Hence the surfactant concentration on the western part of the upper hemisphere of the droplet is higher as compared to the south west part of the lower hemisphere. This has been shown in the surface plot in Fig. 5, where the surfactant distribution is plotted for $\lambda = 0.5$, $\beta = 0.5$ and $k = 3$. It is quite complicated to show the variation of surfactant concentration on the deformed surface of a droplet due to the presence of the surface divergence vector. However if we project the surfactant distribution to an undeformed spherical droplet as $\tilde{\Gamma} = \Gamma \left( r_s^2 / \mathbf{n} \cdot \mathbf{r} \right)$ the surface divergence vector can be evaluated on a sphere.[38] Due to a higher surfactant concentration on the north-western portion of the droplet, the surface tension at that region of the droplet is low. Hence a net gradient in surface tension in the transverse direction is present which the drives the droplet in the cross stream direction. In the present case, as shown in Fig. 5, there is a higher surface tension in the south-eastern portion of the droplet as compared to the north-western portion and hence the droplet migrates towards the flow centerline. This is the 'Regime 1' in the variation of the cross-stream migration velocity as shown in Fig. 4(b). However, depending on the value of $\lambda$, $k$ and $\beta$, the surfactant distribution gets altered such that the direction of transverse migration of the droplet may change as shown in 'Regime 2' of the above figure.

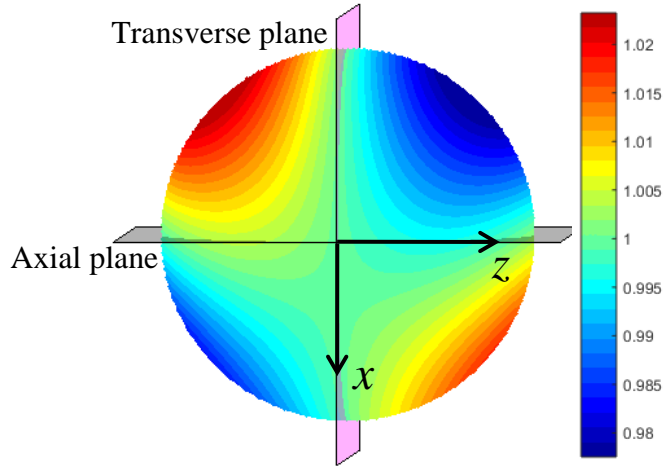

Fig. 5. Surface plot showing the distribution of the surfactants along the droplet surface for $\beta = 0.5$ and $k = 3$. The other parameters involved in the above plot are $R = 5$, $e = 1$, $\lambda = 0.5$ and $Ca = 0.1$.

Figure 6 supports the nature of variation of the cross-stream droplet migration velocity. In this figure the surfactant distribution on both transverse as well as axial planes of the droplet are shown. Figure 6(a) shows the variation of surfactant concentration with the azimuthal angle in two different planes $(\theta = \pi/4, 3\pi/4)$ on either side of the axial plane $(\theta = 0)$. As can be seen from this figure there is an asymmetry in the surfactant distribution about the axial plane when the droplet is placed at an off center position. This is the reason for the lateral migration of the droplet towards the flow centerline. The value of the different parameters used in this figure are $k = 1$, $\lambda = 1$, $\beta = 0.5$, $e = 1$, and $R = 5$. In Fig. 6(b), $k$ has been increased to 3. As a result, the maximum value of surfactant concentration $(\Gamma_{max})$ at the $\theta = \pi/4$ plane increases and the minimum value $(\Gamma_{min})$ at $\theta = 3\pi/4$ plane decreases because of increased surfactant transport due to convection along the droplet surface. Thus with increase in $k$, the magnitude of $|\Gamma_{max} - \Gamma_{min}|$ increases, which causes a higher surface tension gradient and hence a larger surfactant Marangoni stress acting in the transverse direction. Hence the cross-stream migration velocity decreases. Fig. 6(c) and 6(d) show the surfactant distribution in two planes $(\varphi = \pi/4, 3\pi/4)$ on either side of the transverse plane $(\varphi = 0)$ for two different values of $k$. In each of the plane we can see from Fig. 6(c) and 6(d) there is an asymmetry in the surfactant distribution that cause the axial migration of the droplet.

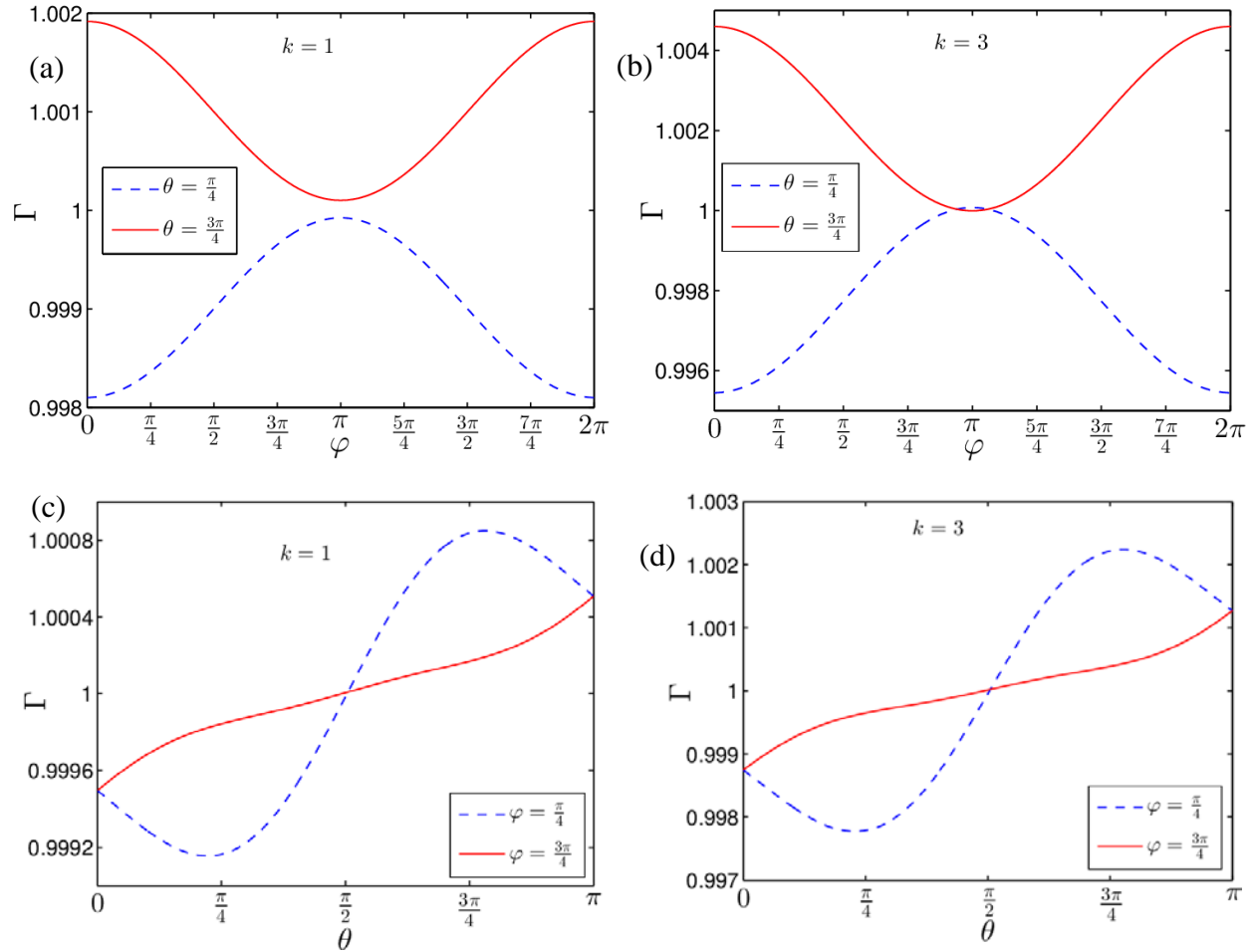

Fig. 6. Variation of surfactant concentration with the azimuthal angle along two axial planes $(\theta = \pi/4, 3\pi/4)$ for (a) $k =1$ and (b) $k = 3$. Surfactant distribution plotted with respect to polar angle along two transverse planes $(\varphi = \pi/4, 3\pi/4)$ for (c) $k =1$ and (d) $k = 3$. The other parameters used in this plot are $R = 5$, $e =1$, $\lambda =1$, $\beta = 0.5$ and $Ca = 0.1$.

## 2. Effect of the elasticity parameter, $\beta$

The effect of $\beta$ on the migration velocity of the droplet is next investigated. This time we keep the property parameter a constant and vary $\beta$. Referring back to Fig. 2(a) and 2(b), we see in each of the figures that increase in $\beta$ $(\beta = 0, 0.25, 0.5)$ leads to a decrease in the axial migration velocity of the droplet. This decrease in the magnitude of the axial velocity is higher for a larger value of $k$. The 'blue solid line', indicates the variation of axial velocity for a clean droplet. The droplet migration velocity is the largest for this case. The effect of $\beta$ is more prominent for a low viscous droplet as compared to a highly viscous one.

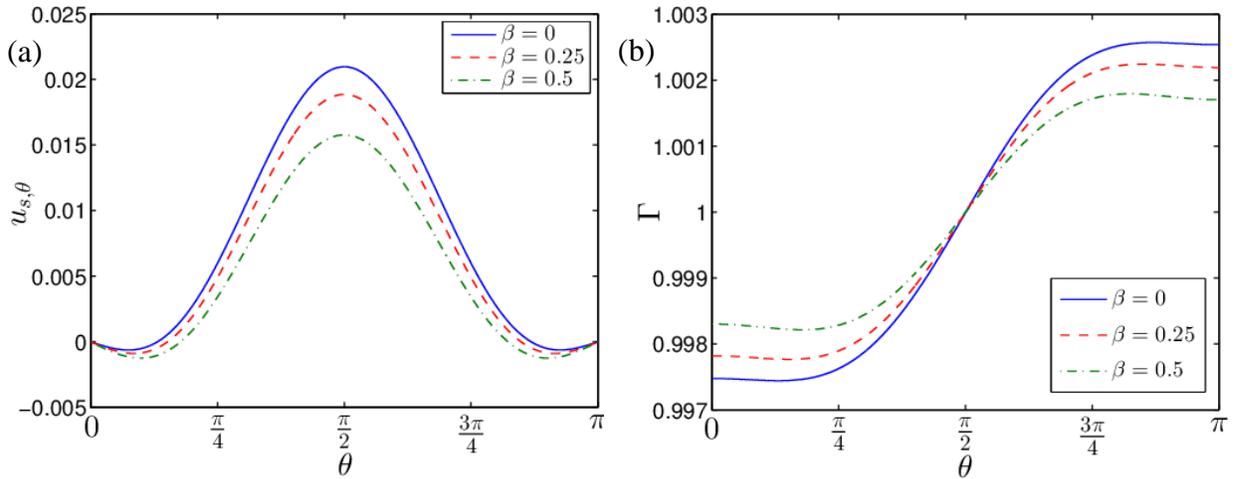

Fig. 7. (a) Variation of surface velocity with $\theta$ along a transverse plane, $\varphi = \pi/2$, and for $\beta = 0, 0.25, 0.5$. (b) Variation of surfactant concentration with the polar angle $(\theta)$ about a transverse plane $(\varphi = \pi/2)$ for different values of $\beta (= 0, 0.25, 0.5)$. The other parameters used in this plot are $R = 5$, $e = 1$, $\lambda = 1$, $k = 2$ and $Ca = 0.1$.

A physical reasoning for such effect of $\beta$ on the droplet dynamics can be provided by considering the significance of $\beta$. Increase in the value of $\beta$, in fact, increases the sensitivity of surface tension towards the surfactant distribution on the droplet surface. That is, for the same surfactant distribution, the region near the rear stagnation point having a higher surfactant concentration has a rather lower surface tension for a higher value of $\beta$. In a similar manner, the region near the front stagnation point has a higher surface tension for a larger value of $\beta$. Thus effectively the surface tension gradient along the droplet surface increases for a higher $\beta$, resulting in an increase in the surfactant induced Marangoni stress opposing the motion of the droplet. Hence the axial migration velocity of the droplet decreases (Fig. 2). Due to the reduction in axial velocity, the surface velocity of the droplet also reduces with increase in $\beta$ (Fig. 7(a)), leading to a decrease in surfactant concentration gradient about the droplet surface $(|\Gamma_{max} - \Gamma_{min}|)$ as shown in Fig. 7(b).

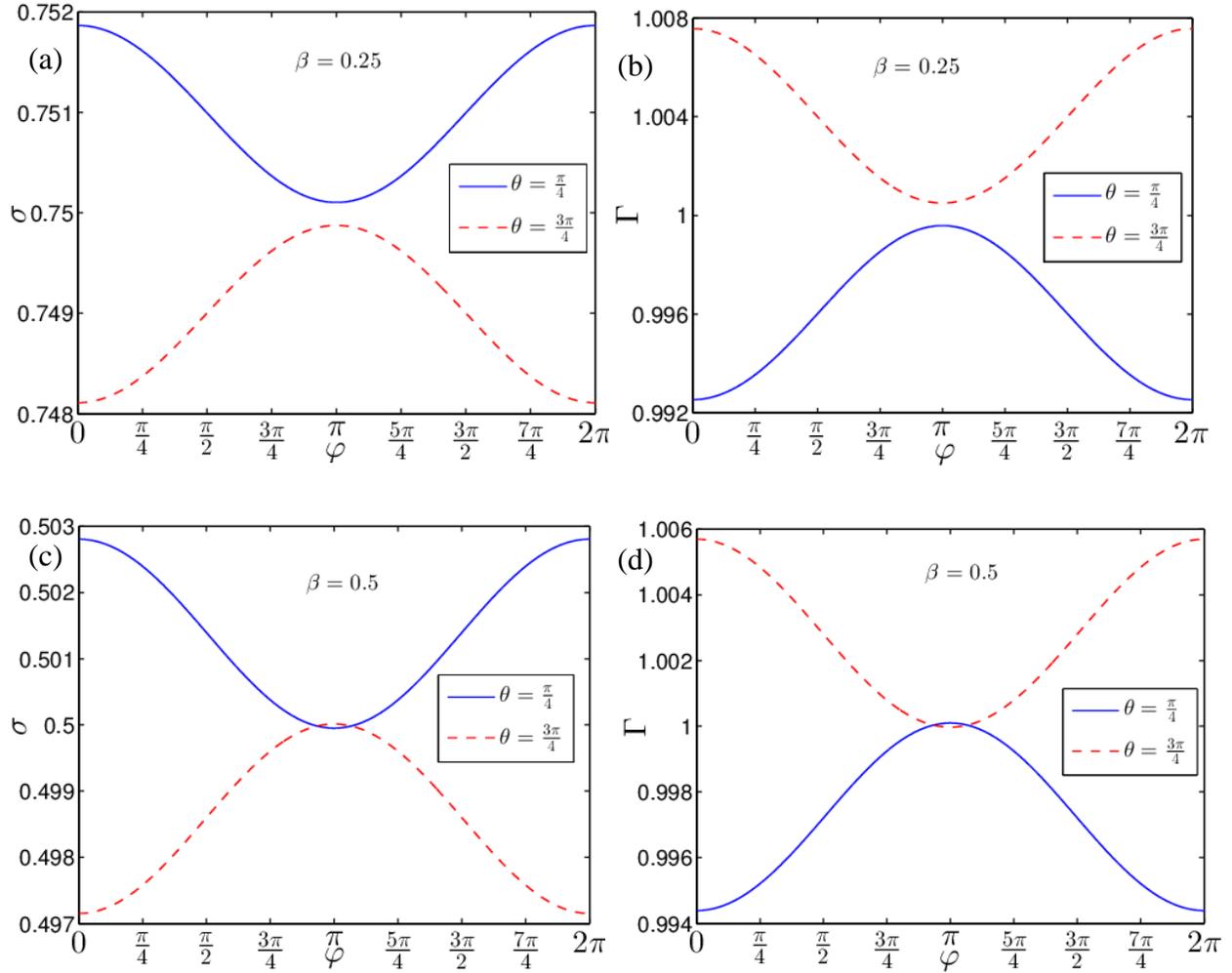

Fig. 8. Variation of surface tension as well as surfactant concentration along two axial planes $(\theta = \pi/4, 3\pi/4)$. Fig. (a) and Fig. (c) shows the variation of $\sigma$ for $\beta = 0.25$ and $\beta = 0.5$ respectively. Fig. (b) and Fig. (d) shows the variation of $\Gamma$ for $\beta = 0.25$ and $\beta = 0.5$ respectively. The other parameters are $\lambda = 0.5,\ k = 3,\ Ca = 0.1,\ e = 1$ and $R = 5$.

Now we look into the effect of $\beta$ on the cross-stream velocity of the droplet for a fixed value of $k$. We again refer to Fig. 4(a) where we see the variation of the cross-stream velocity of the droplet with $\lambda$ for different value of $\beta (= 0, 0.25, 0.5)$. We see that for low viscous droplets $(\lambda \leq 0.75)$, increase in $\beta$ reduces the droplet migration velocity, whereas for higher viscous droplets $(10.5 \geq \lambda \geq 0.75)$ larger $\beta$ results in an increase in the lateral migration velocity of the droplet or a change in the direction of cross-stream migration of the droplet altogether, provided $k$ is kept constant. For even higher viscous droplets $(\lambda \geq 11)$, there is hardly any influence of the surfactant distribution on the droplet migration as it effectively behaves as a particle and hence always migrates towards the flow centerline. In Fig. 4(b), we see that for $\lambda =1$, $k=3$ and $\beta =0.5$, a droplet migrates to the flow centerline. However, when the value of $\beta$ is reduced to 0.25 the same droplet starts migrating away from the flow centerline.

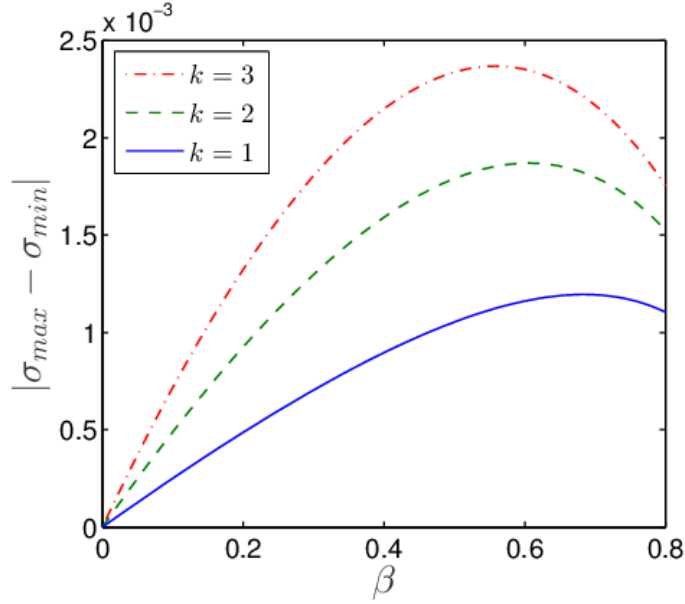

Fig. 9. Variation of the surface tension gradient along the transverse plane $(\varphi = \pi/2)$ with $\beta$ for three different values of $k(=1,2,3)$. The other parameters are $\lambda = 0.5$, $Ca = 0.1$, $e=1$ and $R=5$.

A physical reasoning for the nature of variation of the cross-stream migration velocity can be given if we look into the variation of surfactant concentration and associated variation in surface tension across the axial plane of the droplet. In Fig. 8(b) we have shown the variation of surfactant concentration along two axial planes $(\theta = \pi/4, 3\pi/4)$ for $\lambda = 0.5$, $\beta = 0.25$ and $k=3$. The corresponding variation in surface tension is shown in Fig. 8(a). Clearly, it can be seen that there is an asymmetry in the variation in surface tension in either of the axial planes which is the reason for the transverse migration of the drop. Increase in the value of $\beta$ from 0.25 to 0.5 results in an increase in the sensitivity of surface tension towards the surfactant distribution along the droplet surface. As a result the surface tension, for the case of $\beta = 0.5$, in a region of higher (lower) surfactant concentration decreases (increases) as compared the case when $\beta = 0.25$. It is seen on comparison of Fig. 8(a) with Fig. 8(c) that the surface tension gradient $(|\sigma_{max} - \sigma_{min}|)$ increases. This increase in surface tension in between the two axial planes result in an increase of the Marangoni stress which acts against the cross-stream motion of the droplet, that is in the direction away from the flow centerline. As a result the cross-stream migration velocity reduces with increase in $\beta$, which can be seen in the 'Regime 1' of Fig. 4.

Due to the decrease in the migration velocity, the surface velocity of the droplet also reduces which results in reduced surfactant transport. Hence the surfactant concentration gradient $\left(\left|\Gamma_{max} - \Gamma_{min}\right|\right)$ along either of the axial planes, $\theta = \pi/4, 3\pi/4$, reduces with increase in $\beta$. This can be noticed on comparison of Fig. 8(d) with Fig. 8(b). For the other regimes, a similar reasoning can be given to support the nature of variation of the cross-stream migration velocity with $\lambda$.

The driving force resulting in the migration of the droplet in the flow field originates from the variation of the surface tension along the droplet surface. To get a further insight regarding the transverse migration of the droplet, we plot the variation in surface tension gradient $\left(\left|\sigma_{max} - \sigma_{min}\right|\right)$ with $\beta$ for different values of $k$ in Fig. 9. This surface tension gradient is obtained by taking the difference of the maximum and the minimum surface tension along the transverse plane, $\varphi = \pi/2$. We see that, in general, increase in both $\beta$ (for a fixed value of $k$) and $k$ (for a fixed value of $\beta$) increases the surface tension gradient and hence the Marangoni stress. However, if the value of $\beta$ is increased largely $(\beta > 0.6)$, reduction in surface tension gradient is seen to occur that causes a decrease in the Marangoni stress and hence an increase in the cross-stream migration velocity. Thus there is a 'peak' formation in Fig. 9 which denotes the magnitude of the maximum value of the surface tension gradient. On the right (left) side of the 'peak' increase in $\beta$ decreases (increases) the surface tension gradient, provided $k$ is kept constant. However, increase in the value of $k$ shifts the 'peak' towards the left and the increase of surface tension gradient with $\beta$ becomes steeper. We in our analysis, for the limiting case of $Pe_s \ll 1$, have considered $\beta \leq 0.5$.

We now highlight the effect of both the parameters $k$ and $\beta$ on the cross-stream migration velocity of the drop with the help of a contour plot as shown in Fig. 10. If we look carefully into Fig. 4 (a) and Fig. 4(b), we can clearly see that there are three regimes of variation of the cross-stream migration velocity. These three regimes have been shown in the contour plots in Fig. 10. In the first regime, for a low viscous droplet $(\lambda = 0.1)$, increase in $k$ greatly reduces the lateral migration velocity (Fig. 10(a)). The parameter $\beta$, for such a low viscous drop, has also the same effect on the cross-stream migration velocity of the droplet, however, the magnitude of the decrease in the cross-stream droplet velocity with increase in $\beta$ is far less as compared to the effect of $k$. In this regime, no change in the direction of lateral migration of the droplet occurs and it thus always move towards the centerline of flow. For the second regime with a moderate value of $\lambda$, say $\lambda = 1$ (Fig. 10(b)), for low values of both $\beta$ and $k$, the droplet migrates away from the flow centerline. With increase in both the parameter values the cross-stream velocity of the droplet reduces until it reaches a point where for a particular pair of values of $\beta$ and $k$, there occurs no cross stream migration velocity of the droplet. On further increase in values of $\beta$ and $k$, the cross-stream migration velocity of the droplet, which now migrates towards the centerline of flow. The third regime considers the case of a highly viscous droplet $(\lambda \gg 1)$ and is shown in Fig 10(c). Here increase in both $\beta$ and $k$ increases the cross-stream migration velocity of the droplet, although the increase in the magnitude of the velocity is small. For this regime there is no change in the direction of migration of the droplet and the droplet migrates towards the flow centerline.

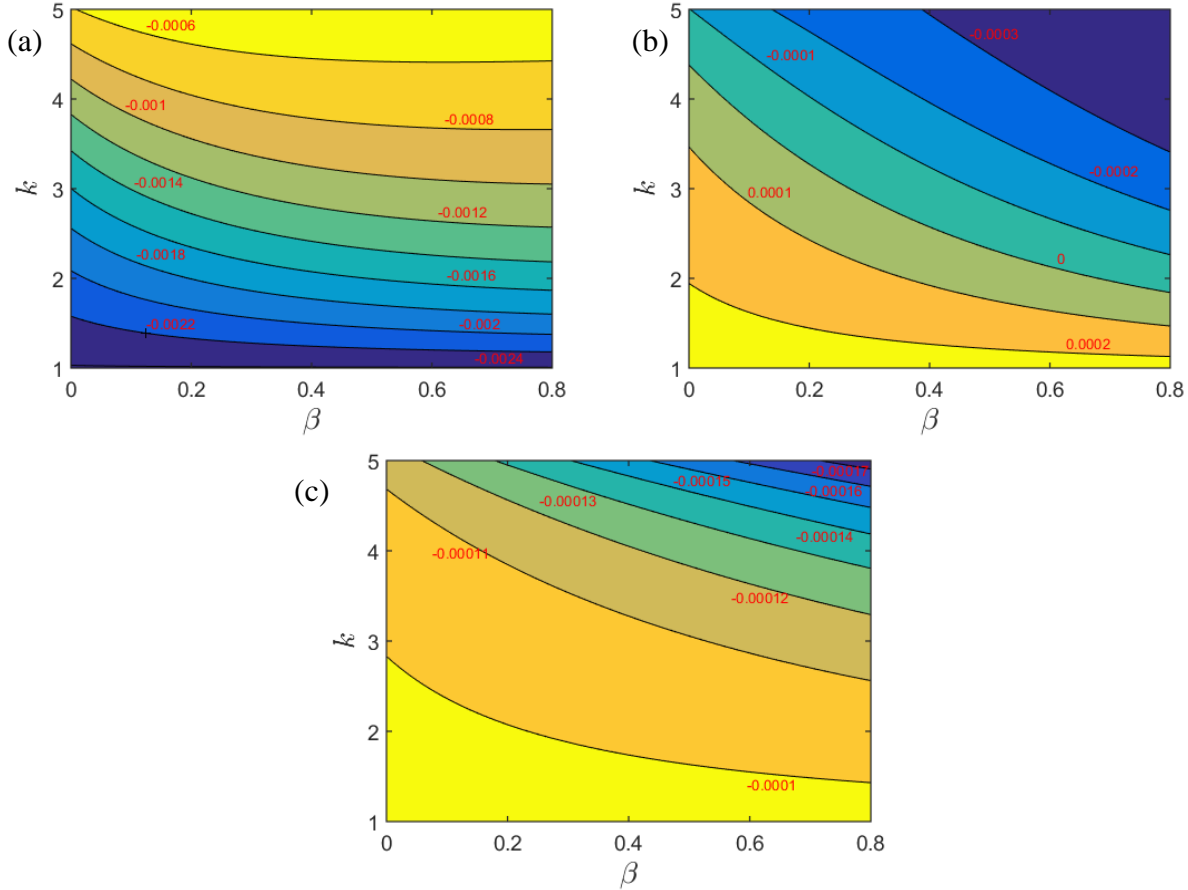

Fig. 10. Contour plot showing the variation of cross-stream migration velocity for the three regimes. Fig. (a) shows the first regime for $\lambda = 0.1$, Fig. (b) shows the second regime for $\lambda = 1$ and Fig. (c) shows the third regime where $\lambda = 20$. The other parameters involved in the above plot are $R = 5$, $e = 1$ and $Ca = 0.1$.

The above variation of cross-stream migration velocity is due to both shape deformation as well as the tangential Marangoni stress on the droplet interface. The tangential Marangoni stress is represented by $\{\beta/(1-\beta)Ca\}\nabla_s \Gamma$. The non-uniform distribution of surfactants on the surface of the droplet, due to the imposed flow, is responsible for either of the effects: droplet deformation and generation of Marangoni stress. As the surfactant distribution on the surface of the droplet is affected by both $\beta$ and $k$, either of them are also responsible for affecting the droplet deformation as well as the Marangoni stress.

We next look into the effect of both $\beta$ and $k$ on the temporal variation of the lateral position of the droplet. We obtain the lateral migration of the droplet as function of time by substitution of $U_x = de/dt$ in the expression of the x-component of velocity as given in equation (41). The solution for the lateral position of the droplet, $e$, as a function of time, $t$ is given below

$$\left. \begin{array}{l} e(t) = e_0 \exp\left(-\dfrac{t}{t_c}\right), \\ \text{where,} \\ t_c = -\dfrac{1}{Ca}\left(\dfrac{c_6}{c_1\beta^4 + c_2\beta^3 + c_3\beta^2 + c_4\beta + c_5}\right). \end{array} \right\} \quad (43)$$

Here $e_0 = e(t=0)$ is the initial position of the droplet and $t_c$ is the characteristic time constant. When $t_c > 0$, there is an exponential decrease in the distance between the initial position of the droplet and the flow centerline. The larger the magnitude of $t_c$, the quicker the droplet reaches its steady state position.

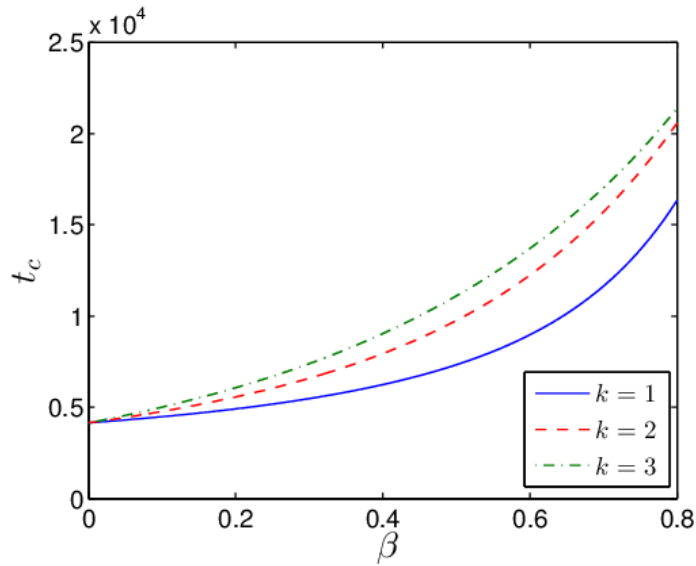

Fig. 11. Variation of time constant $t_c$ with $\beta$ for different values of $k$. The values of the other parameters are $R = 5$, $\lambda = 0.1$ and $Ca = 0.1$.

For $\lambda = 0.1$, higher $\beta$ values results in higher values of $t_c$. This can be seen from Fig. 11 where we have shown the variation of the characteristic time constant with $\beta$ for different values of $k$. It is seen from this figure that increase in $\beta$ for a constant value of $k$, or increase in $k$ for a constant value of $\beta$ increases $t_c$. Hence the time required for the drop to reach its steady state position, in either of the cases, decreases.

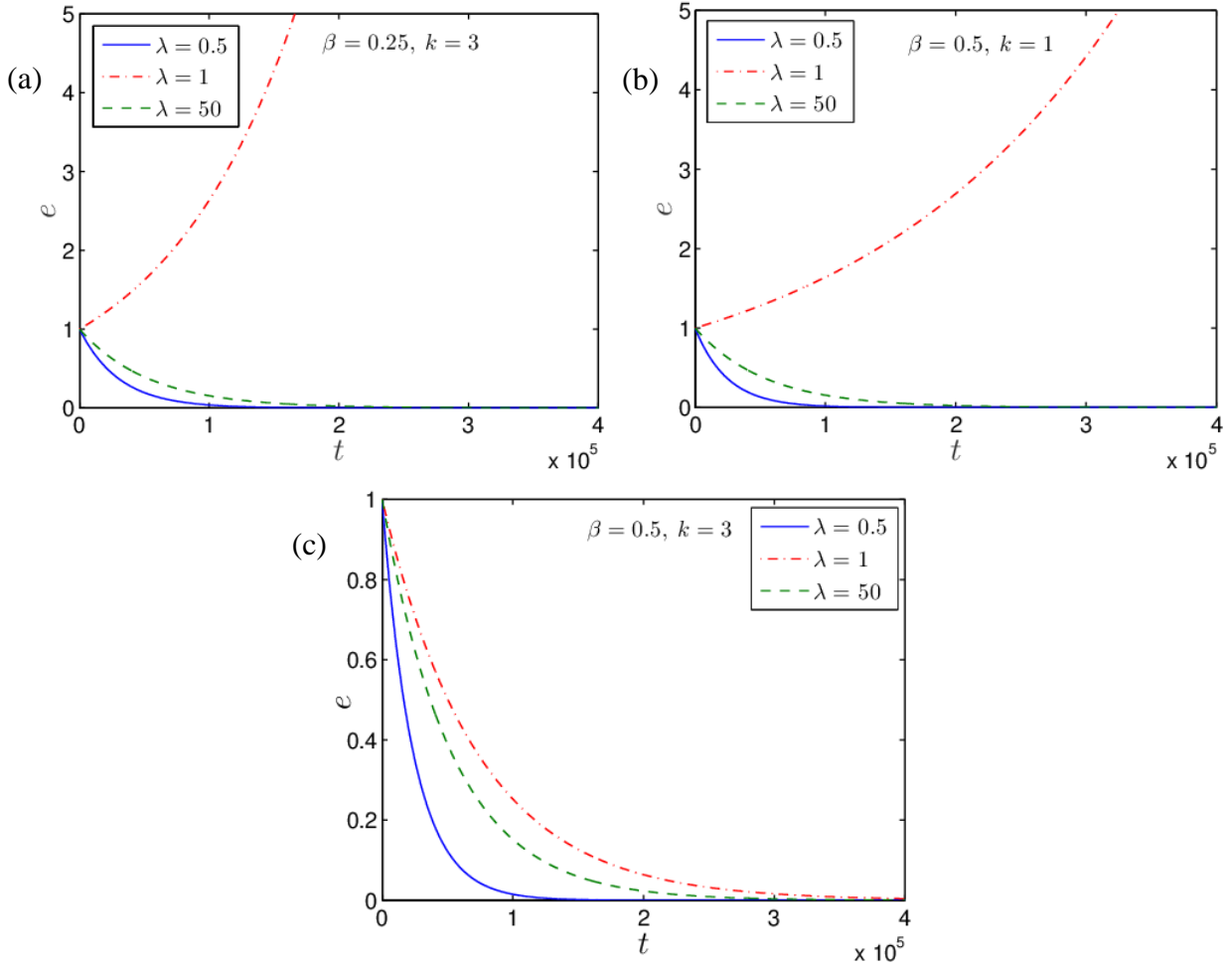

Fig. 12. Temporal variation of the lateral migration of the droplet is shown for three different cases of varying $k$ and $\beta$ namely (a) $k = 3, \beta = 0.25$, (b) $k = 1, \beta = 0.5$ and (c) $k = 3, \beta = 0.5$. Each of the above plots are shown for three separate values of $\lambda(= 0.5, 5, 50)$. The other parameters are $R = 5$ and $Ca = 0.1$. The initial position of the droplet is taken as $e(t = 0) = 1$.

Comparing Fig. 12(a) with 12(c), we see that for a system with $\beta = 0.5$, the time taken by a low viscous droplet $(\lambda = 0.5)$ to reach its steady state position is less as compared to the case for which $\beta = 0.25$, provided $k$ is kept constant. Also we can see from comparison between Fig. 12(b) and 12(c) that for a constant value of $\beta(=0.5)$, increase in the value of $k$ reduces the time taken by the droplet to reach its steady state position, when $\lambda = 0.5$.

It can also be seen from the figures above that both variation in $\beta$ and $k$ can cause change in direction of lateral migration of the droplet. Comparing Fig. 12(a) with 12(c), we see that for $\lambda = 1$, the droplet, which initially was migrating towards the centerline of flow for $\beta = 0.5$, starts migrating away from the centerline for $\beta = 0.25$. The value of $k$ is kept constant at 3. Again on comparison of Fig. 12(b) and Fig. 12(c), we see that a decrease in the value of $k$ from $k = 3$ to $k = 1$ results in a change in direction of lateral migration of the droplet, provided $\beta$ is kept constant $(\beta = 0.5)$. However, a low viscous droplet $(\lambda = 0.1)$ migrates towards to the centerline irrespective of the value of $k$.

## B. High surface Péclet number Limit

In this limit the droplet migration velocity can be expressed as

$$\mathbf{U} = \left[\left(1 - \frac{2}{3R^2}\right) - \frac{e^2}{R^2}\right]\mathbf{e}_z - \left[Ca\frac{e}{6R^4}\left(\frac{4}{\beta} - 3\right)\right]\mathbf{e}_x. \tag{44}$$

From the above expression it can be said that the axial velocity in this limit is independent of the surfactant distribution. It can be seen that the $O(Ca)$ correction to the droplet migration velocity is always negative. It can also be inferred from the above expression that the leading order solution for droplet migration velocity is the same as that for a rigid spherical particle in a Poiseuille flow. This, however, is not true if $O(Ca)$ correction in droplet migration velocity is taken into consideration. For the cross-stream migration velocity, there is a clear dependence on $\beta$. We first show the effect the elasticity parameter, $\beta$ on the cross-stream migration velocity of the droplet.

### 1. Effect of the elasticity parameter, $\beta$

As can be seen from the expression of cross-stream component of the migration velocity in equation (44), there is no dependence of the same on $\lambda$. Thus we show the variation of the cross-stream migration velocity with $\beta$ in Fig. 13. As seen from Fig. 13, the cross-stream migration velocity of the droplet decreases rapidly with increase in $\beta$. The cross-stream velocity vanishes asymptotically with further increase in $\beta$.

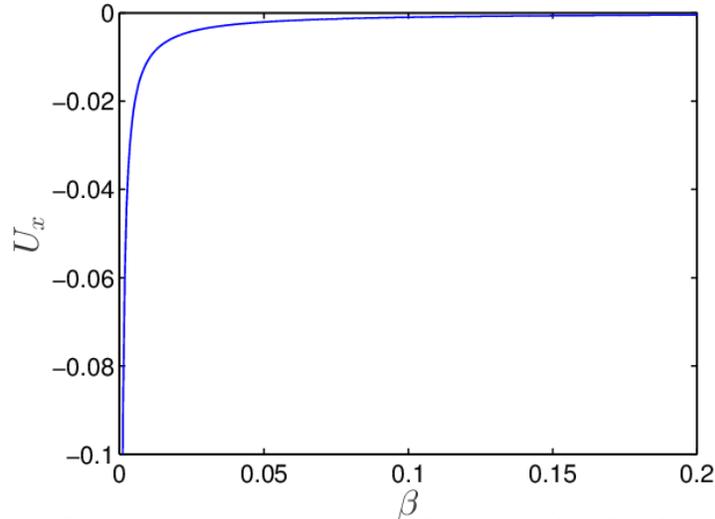

Fig. 13. Variation of cross-stream migration velocity of the droplet with $\beta$. The different parameters used for this plot are $R = 5$, $e = 1$, and $Ca = 0.1$.

An explanation can be given if we look into the surface plot for the surfactant distribution over the droplet surface (Fig. 14). Fig. 14 shows the distribution of surfactants by projecting it on an undeformed spherical droplet surface using the transformation $\tilde{\Gamma} = \Gamma\left(r_s^2/\mathbf{n}\cdot\mathbf{r}\right)$, where $\tilde{\Gamma}$ denotes the projected surfactant concentration.

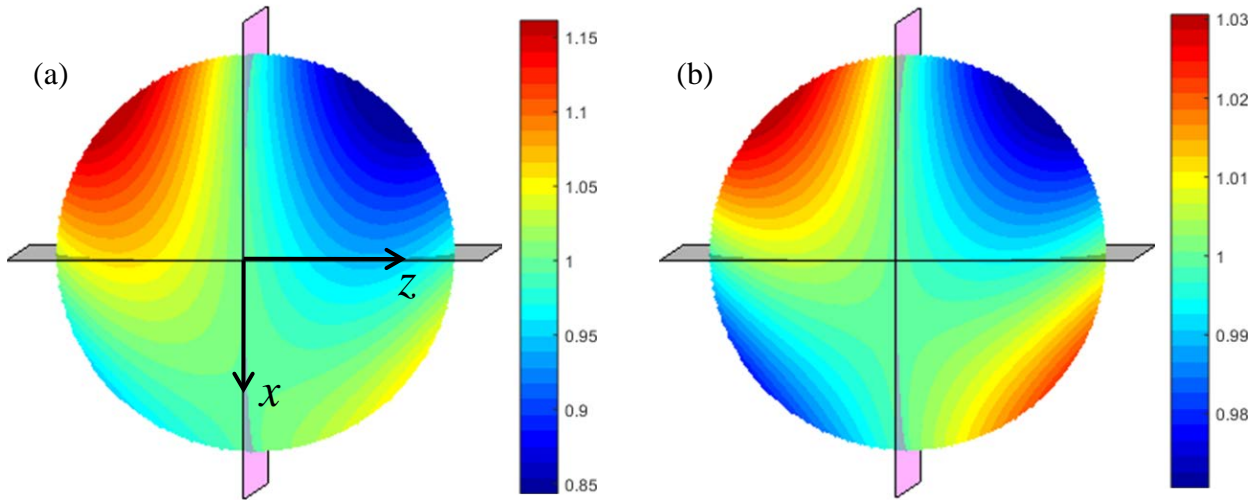

Fig. 14. Surface plot showing the distribution of the surfactants along the droplet surface for (a) $\beta = 0.1$, (b) $\beta = 0.6$ (c). The other parameters involved in the above plot are $R = 5$, $e = 1$, $\lambda = 0.1$, $k = 50$ and $Ca = 0.1$.

It is seen that increase in $\beta$ increases the sensitivity of surface tension towards the surfactant distribution along the droplet surface. Thus for the same non-uniform distribution of surfactants along the droplet surface, there is a higher gradient in surface tension for a higher value of $\beta$. Hence a higher value of $\beta$ ensures a larger value of Marangoni stress, which prevents the droplet from migrating towards the flow centerline and thus reduces the migration velocity of the droplet. This in turn results in a decrease in the surface velocity of the droplet and hence a reduction in the non-uniformity in surfactant distribution. In other words, magnitude of $|\Gamma_{max} - \Gamma_{min}|$ as can be seen on comparison between Fig. 14(a) and Fig. 14(b), reduces with increase in $\beta$.

Now we look into the time variation of the lateral position of the droplet for different values of $\beta (= 0.1, 0.3, 0.6)$. The other parameters used are $e = 1, R = 5$. We proceed in a manner as was done for the limiting case of $Pe_s \ll 1$. We obtain an expression for the transverse position of the droplet as a function of time by substituting $U_x = de/dt$ in the expression for the x-component of droplet migration velocity. On integration of the same equation, we obtain the lateral position of the droplet as.

$$
\left.
\begin{aligned}
e(t) &= e_0 \exp\left(-\frac{t}{t_c}\right), \\
\text{where,} & \\
t_c &= \frac{Ca}{6R^4}\left(\frac{4}{\beta} - 3\right).
\end{aligned}
\right\} \quad (45)
$$

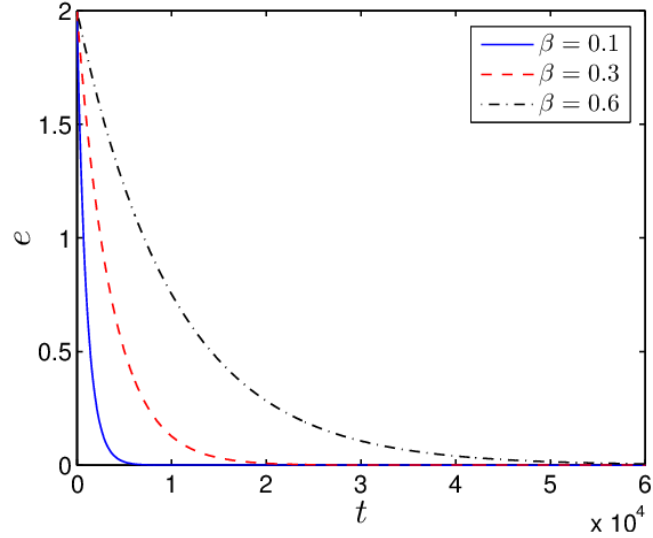

Fig. 15. Variation of the transverse position of the droplet with time for different values of $\beta\,(0.1, 0.3, 0.6)$. The other parameters involved are $R = 5$, $k = 50$ and $Ca = 0.1$.

From the equation (45) it can be seen that a decrease in the value of $\beta$ increases $t_c$ and thus the time required for the droplet to reach its steady state position decreases. This is exactly seen in Fig. 15, that is, the time taken by the droplet to reach the flow centerline is larger for a higher value of $\beta$. This is expected as the cross-stream migration velocity reduces with increase in $\beta$. It is also seen from Fig. 15 that the droplet always migrates towards the centerline of flow.

## C. Comparison with the results of Pak et al. (2014) and Hanna and Vlahovska (2010)

As was mentioned in the introduction, Pak et al as well as Hanna and Vlahovska had found out the droplet migration velocities for the two different limiting conditions namely: (i) $Pe_s \ll 1$ and (ii) $Pe_s \gg 1$, respectively without considering the effect of shape deformation. We, in our present problem, however, have considered the effect of small shape deformation. In order to show the effect of the same on the cross-stream migration velocity of the droplet, we compare the result obtained in our work with each of theirs. For the low surface Péclet number limit, we see that the cross stream migration velocity of a deformable surfactant laden droplet is almost 3 times higher than that obtained by Pak et al for a spherical surfactant covered droplet.[33] The different parameters used for the purpose of this comparison are: $Pe_s = 0.1$, $k = 1$, $\beta = 0.5$, $\lambda = 1$, $e = 1$ and $R = 5$. For the other limiting case of $Pe_s \gg 1$, the magnitude of the lateral migration velocity is higher than that obtained by Hanna and Vlahovska.[32] The deformation of a surfactant laden droplet effectively enhances its cross-stream migration velocity.

## V. CONCLUSION

In this paper, we have investigated the droplet motion in the presence of three effects, namely presence of a bulk flow, shape deformation of the droplet and Marangoni stress induced by the non-uniform distribution of the surfactants. Assuming negligible inertial effects and thermal convection, we have asymptotically solved for the surfactant concentration as well as the flow field for the following two limits: (a) surfactant transport dominated by surface diffusion that is $Pe_s \ll 1$, (b) surface convection dominated surfactant transport or $Pe_s \gg 1$. Analytical expressions for the droplet velocity, both axial and transverse components, are obtained for either of the above mentioned limits. Some of the important points in the present analysis are mentioned below

(i) For small $Pe_s$, increase in either $\beta$ or $k$, reduces the axial velocity of the droplet, the decrease in the velocity being larger for the case of low viscous droplet.

(ii) For small $Pe_s$, cross-stream migration of the droplet can be broadly divided into three regimes. In the first regime $(\sim \lambda \leq 0.7)$, increase in both $\beta$ and $k$ decreases the droplet velocity but the droplet always migrates towards the channel centerline without any change in direction. In the second regime $(\sim 0.75 \leq \lambda \leq 10.5)$, there is an increase in the cross-stream migration velocity of the droplet with an increase in either $\beta$ or $k$. In this regime reversal in the direction of droplet migration may take place depending on the values of $\beta$ and $k$, that is the droplet may migrate away from the centerline of flow. In the third regime $(\sim \lambda \geq 11)$, the droplet always migrates towards the centerline of the flow. Change in either $\beta$ or $k$ hardly has any effect on the cross-stream velocity of the droplet.

(iii) Under the limit of small $Pe_s$, it is seen that for $\lambda \sim O(1)$, the droplet starts migrating away from the flow centerline if either of the parameters, $\beta$ or $k$, is decreased. For any other value $\lambda$, the time required for the droplet to reach a steady state position decreases with increase in either $k$ or $\beta$.

(iv) For high $Pe_s$, the axial velocity is found to be independent of the surfactant distribution. However, the cross-stream migration velocity is dependent directly on the elasticity parameter $\beta$, which when increased results in a rapid fall of the lateral migration velocity. The droplet, in this limiting condition, always moves towards the centerline of flow without any change in its direction.

# APPENDIX A: EXPRESSION OF THE CONSTANT COEFFICIENTS PRESENT IN EQUATIONS (25), (28), (33), (30) AND (31)

The constant coefficients present in the expression of the leading order surfactant concentration for the low $Pe_s$ limiting condition, given in equation (25), are written below

$$\begin{aligned}\Gamma^{(0)}_{1,0} &= \frac{2k}{R^2}\left\{\frac{(1-\beta)}{(3\lambda+2-k)\beta-3\lambda-2}\right\},\\ \Gamma^{(0)}_{3,0} &= \frac{7k}{6R^2}\left\{\frac{(1-\beta)}{(k-7\lambda-7)\beta+7+7\lambda}\right\},\\ \Gamma^{(0)}_{2,1} &= \frac{5ke}{3R^2}\left\{\frac{(1-\beta)}{(5\lambda+5-k)\beta-5-5\lambda}\right\}.\end{aligned} \quad (A1)$$

The constant coefficients present in the expression for $O(Ca)$ correction to the droplet shape for the low surface Péclet limit is given below

$$\begin{aligned}L^{(Ca)}_{2,1} &= -\frac{5e}{12R^2}\left\{\frac{(19\lambda+16)(1-\beta)+4k\beta}{5(1-\beta)(1+\lambda)+k\beta}\right\},\\ L^{(Ca)}_{3,0} &= \frac{7}{60R^2}\left\{\frac{3(10+11\lambda)(1-\beta)+5k\beta}{7(1+\lambda)(1-\beta)+k\beta}\right\}.\end{aligned} \quad (A2)$$

The expressions for the constants in $O(Ca)$ surfactant concentration as shown in equation (31) are given below

(i) For $i=0, j=0$

$$\Gamma^{(Ca)}_{0,0} = -\frac{6}{5}\Gamma^{(0)}_{2,1}L^{(Ca)}_{2,1} - \frac{2}{7}\Gamma^{(0)}_{3,0}L^{(Ca)}_{3,0} - \frac{6}{5}\left\{L^{(Ca)}_{2,1}\right\}^2 - \frac{5}{7}\left\{L^{(Ca)}_{3,0}\right\}^2 \quad (A3)$$

(ii) For $i=1, j=1$

$$a_{1,1}^{(Ca)} = 0,$$

$$b_{1,1}^{(Ca)} = 0,$$

$$c_{1,1}^{(Ca)} = 0,$$

$$d_{1,1}^{(Ca)} = -5e\beta^2(1-\beta)\{12\lambda(1-\beta)+7\beta+12\},$$

$$f_{1,1}^{(Ca)} = -e\beta(1-\beta)^2\{720(1-\beta)\lambda^2+(602\beta+1290)\lambda+135\beta+570\},$$

$$h_{1,1}^{(Ca)} = -e(1-\beta)^3\begin{Bmatrix}2100\lambda^3(1-\beta)+(6933\beta+5370)\lambda^2\\+(6541\beta+4470)\lambda-1400\beta+1200\end{Bmatrix},$$

$$l_{1,1}^{(Ca)} = -3e(1-\beta)^4(8462\lambda^3+12885\lambda^2+2663\lambda-1820),$$

$$m_{1,1}^{(Ca)} = 0,$$

$$q_{1,1}^{(Ca)} = 20\begin{bmatrix}\{\beta k+(1-\beta)(3\lambda+2)\}\{\beta k+5(1-\beta)(1+\lambda)\}\\\{\beta k+7(1-\beta)(1+\lambda)\}\{\beta k+3(1-\beta)(1+\lambda)\}\end{bmatrix}R^4, \quad (A4)$$

(iii) For $i=2, j=0$

$$a_{2,0}^{(Ca)} = 900\beta^5(1-\beta)\{35e^2(1+\lambda)+7\beta-2\beta e^2-28\beta\lambda e^2\},$$

$$b_{2,0}^{(Ca)} = -5bt^4(1-\beta)^2\begin{pmatrix}-27020\beta+97065\beta\lambda e^2-20880\beta e^2-40992\beta\lambda\\+120015\beta\lambda^2 e^2-149400e^2-306900\lambda e^2-3500-157500\lambda^2 e^2-3500\lambda\end{pmatrix},$$

$$c_{2,0}^{(Ca)} = \beta^3(1-\beta)^3\begin{pmatrix}3411520\beta\lambda+3058200\beta e^2+2706648\beta\lambda^2+1258650\beta\lambda e^2+784700\beta-\\7239825\beta\lambda^2 e^2-5350275\beta\lambda^3 e^2+560000+7623000\lambda^3 e^2+21933000\lambda^2 e^2\\+20983500\lambda e^2+1092000\lambda+6673500e^2+532000\lambda^2\end{pmatrix},$$

$$d_{2,0}^{(Ca)} = -\beta^2(1-\beta)^4\begin{pmatrix}-14275800\beta\lambda-28713600\beta e^2-19035744\beta\lambda^3-34716136\beta\lambda^2\\+1409100\beta+21234150\beta\lambda^4 e^2+29447550\beta\lambda^3 e^2-35273025\beta\lambda^2 e^2\\-72082575\beta\lambda e^2-27558000e^2-189724500\lambda^2 e^2-6247500-5727400\lambda^3\\-118390500\lambda e^2-18217500\lambda-35595000\lambda^4 e^2-134487000\lambda^3 e^2\\-17697400\lambda^2\end{pmatrix},$$

$$f_{2,0}^{(Ca)} = -5\beta(1-\beta)^5 \begin{pmatrix} -15660204\beta\lambda^4 + 3516240\beta\lambda - 24089760\beta e^2 - 38267376\beta\lambda^3 \\ -24339980\beta\lambda^2 + 5251400\beta + 6043590\beta e^2\lambda^5 - 6812370\beta\lambda^4 e^2 \\ -83150595\beta\lambda^3 e^2 - 145375695\beta\lambda^2 e^2 - 99178920\beta\lambda e^2 - 5845000 \\ -16008300 e^2\lambda^5 - 57178800\lambda e^2 - 74346300\lambda^4 e^2 - 137232900\lambda^3 e^2 \\ -125766900\lambda^2 e^2 - 22750000\lambda - 33189800\lambda^2 - 10306800 e^2 \\ -5224800\lambda^4 - 21509600\lambda^3 \end{pmatrix}, \quad (A5)$$

$$h_{2,0}^{(Ca)} = 175(1+\lambda)(1-\beta)^6 \begin{pmatrix} 280000 + 1085000\lambda - 392000\beta + 1575700\lambda^2 + 201600 e^2 \\ -485240\beta\lambda + 1016400\lambda^3 + 1184400\lambda e^2 + 1259916\beta\lambda^2 \\ +1293696\beta e^2 + 245700\lambda^4 + 2746800\lambda^2 e^2 + 2418276\beta\lambda^3 \\ +5960772\beta\lambda e^2 + 3143700\lambda^3 e^2 + 1064448\beta\lambda^4 + 10174392\beta\lambda^2 e^2 \\ +1776600\lambda^4 e^2 + 7705107\beta\lambda^3 e^2 + 396900 e^2\lambda^5 + 2311542\beta\lambda^4 e^2 \\ +116991\beta e^2\lambda^5 \end{pmatrix},$$

$$l_{2,0}^{(Ca)} = 18375(1-\beta)^7 (3\lambda+2)(1+\lambda)^2 \begin{pmatrix} 1197\lambda^4 e^2 + 3696\lambda^3 + 8475\lambda^3 e^2 + 18258\lambda^2 e^2 \\ +6352\lambda^2 + 15096\lambda e^2 + 1840\lambda + 4224 e^2 - 800 \end{pmatrix},$$

$$m_{2,0}^{(Ca)} = 0,$$

$$q_{2,0}^{(Ca)} = 7560\{\beta k + (3\lambda+2)(1-\beta)\}^2 \{\beta k + 5(1+\lambda)(1-\beta)\}^3 \{\beta k + 7(1-\beta)(1+\lambda)\}^2 R^4,$$

(iv) For $i = 2, j = 2$

$$a_{2,2}^{(Ca)} = 0,$$
$$b_{2,2}^{(Ca)} = 0,$$
$$c_{2,2}^{(Ca)} = 0,$$
$$d_{2,2}^{(Ca)} = 0,$$
$$f_{2,2}^{(Ca)} = 20 e^2 \beta (1-\beta)(-35 + 22\beta + 28\beta\lambda - 35\lambda), \qquad (A6)$$
$$h_{2,2}^{(Ca)} = 5 e^2 (1-\beta)^2 \begin{Bmatrix} (427\beta - 700)\lambda^2 + \\ (281\beta - 1100)\lambda + 592\beta - 400 \end{Bmatrix},$$
$$l_{2,2}^{(Ca)} = -25 e^2 (1-\beta)^3 (7\lambda - 5)(19\lambda + 16)(\lambda + 4),$$
$$m_{2,2}^{(Ca)} = 0,$$
$$q_{2,2}^{(Ca)} = 1008(-5\beta - 5\beta\lambda + \beta k + 5 + 5\lambda)^3 R^4,$$

(v) For $i=3, j=1$

$$a_{3,1}^{(Ca)} = 0,$$
$$b_{3,1}^{(Ca)} = 0,$$
$$c_{3,1}^{(Ca)} = 0,$$
$$d_{3,1}^{(Ca)} = 420e\beta^2(1-\beta)(25\beta - 6\beta\lambda + 7 + 7\lambda),$$
$$f_{3,1}^{(Ca)} = 2\beta e(1-\beta)^2 \begin{pmatrix} -8694\beta\lambda^2 + 64710\beta + 65777\beta\lambda \\ +40250\lambda + 28490 + 11760\lambda^2 \end{pmatrix},$$
$$h_{3,1}^{(Ca)} = e(1-\beta)^3 \begin{pmatrix} 522720\beta + 687969\beta\lambda^2 - 15624\beta\lambda^3 \\ +1214951\beta\lambda + 303800 + 388080\lambda^2 \\ +647780\lambda + 44100\lambda^3 \end{pmatrix},$$
$$l_{3,1}^{(Ca)} = 21e(1-\beta)^4 (56560 + 1980\lambda^4 + 83609\lambda^3 + 215055\lambda^2 + 190126\lambda),$$
$$m_{3,1}^{(Ca)} = 0,$$
$$q_{3,1}^{(Ca)} = 4320\{\beta k + (3\lambda + 2)(1-\beta)\}\{\beta k + 5(1+\lambda)(1-\beta)\}\{\beta k + 7(1+\lambda)(1-\beta)\}^2 R^4, \quad (A7)$$

(vi) For $i=4, j=0$

$$a_{4,0}^{(Ca)} = 0,$$
$$b_{4,0}^{(Ca)} = -20\beta^5(1-\beta)(91 + 550e^2),$$
$$c_{4,0}^{(Ca)} = -\beta^3(1-\beta)^2 \begin{pmatrix} 33432\beta\lambda + 30590\beta + 270600\beta e^2 \\ +289850\beta\lambda e^2 + 16275 + 16275\lambda + 19800\lambda e^2 \\ +19800e^2 \end{pmatrix},$$
$$d_{4,0}^{(Ca)} = -\beta^2(1-\beta)^3 \begin{pmatrix} 2642200\beta e^2 + 444017\beta\lambda + 5717800\beta\lambda e^2 \\ +252007\beta\lambda^2 + 194530\beta + 3083850\beta\lambda^2 e^2 \\ +285495\lambda^2 + 316800e^2 + 573195\lambda + 287700 \\ +336600\lambda^2 e^2 + 653400\lambda e^2 \end{pmatrix},$$

$$f_{4,0}^{(Ca)} = -\beta(1-\beta)^4 \begin{pmatrix} 3267453\beta\lambda^2 + 2849826\beta\lambda + 12867800\beta e^2 \\ +1241219\beta\lambda^3 + 16661150\beta\lambda^3 e^2 + 46021800\beta\lambda^2 e^2 \\ +42228450\beta\lambda e^2 + 822710\beta + 1656375 + 4851000\lambda e^2 \\ +4947075\lambda + 5128200\lambda^2 e^2 + 4925025\lambda^2 \\ +1801800\lambda^3 e^2 + 1634325\lambda^3 + 1524600 e^2 \end{pmatrix}, \quad (A8)$$

$$h_{4,0}^{(Ca)} = -35(1+\lambda)(1-\beta)^5 \begin{pmatrix} 129719\beta\lambda^3 + 1301410\beta\lambda^3 e^2 + 346778\beta\lambda^2 \\ +3454550\beta\lambda^2 e^2 + 3037540\beta\lambda e^2 + 310021\beta\lambda \\ +884400\beta e^2 + 92710\beta + 89250 + 55440 e^2 \\ +87675\lambda^3 + 264600\lambda^2 + 194040\lambda e^2 + 266175\lambda \\ +83160\lambda^3 e^2 + 221760\lambda^2 e^2 \end{pmatrix},$$

$$l_{4,0}^{(Ca)} = -1225(1+\lambda)^2(1-\beta)^6 \begin{pmatrix} 40128\lambda^3 e^2 + 6501\lambda^3 + 18087\lambda^2 + 100672\lambda^2 e^2 \\ +16944\lambda + 83072\lambda e^2 + 5340 + 22528 e^2 \end{pmatrix},$$

$$m_{4,0}^{(Ca)} = 0,$$

$$q_{4,0}^{(Ca)} = 1540 \begin{bmatrix} \{\beta k + (3\lambda+2)(1-\beta)\}\{\beta k + 5(1+\lambda)(1-\beta)\}^2 \\ \{\beta k + 7(1+\lambda)(1-\beta)\}^2 \{\beta k + 9(1+\lambda)(1-\beta)\} \end{bmatrix} R^4,$$

(vii) For $i = 4, j = 2$

$$a_{4,2}^{(Ca)} = 0,$$
$$b_{4,2}^{(Ca)} = 0,$$
$$c_{4,2}^{(Ca)} = 0,$$
$$d_{4,2}^{(Ca)} = 0,$$
$$f_{4,2}^{(Ca)} = 100 bt^2 e^2 (1-bt), \quad (A9)$$
$$h_{4,2}^{(Ca)} = 5(1-\beta)^2 e^2 \{187\beta\lambda + 172\beta + 36(1+\lambda)\},$$
$$l_{4,2}^{(Ca)} = 160(1+\lambda)(19\lambda+16)(1-\beta)^3 e^2,$$
$$m_{4,2}^{(Ca)} = 0,$$
$$n_{4,2}^{(Ca)} = 168\{\lambda k - 5\beta(1+\lambda) + 5(1+\lambda)\}^2 \{\beta k - 9\beta(1+\lambda) + 9(1+\lambda)\} R^4,$$

The constants present in the expression of the $O(Ca)$ cross-stream migration velocity (equation (33)) for the low surface Péclet number limit is given below

$$c_1 = \begin{Bmatrix} 15k^4 + (-115 - 152\lambda)k^3 + (943\lambda - 240 + 142\lambda^2)k^2 \\ +(-4339\lambda^2 + 4220 + 1512\lambda^3 - 593\lambda)k \\ -9600 + 18870\lambda^2 - 7290\lambda - 1485\lambda^4 + 15255\lambda^3 \end{Bmatrix},$$

$$c_2 = \begin{Bmatrix} -20k^4 + (392\lambda + 305)k^3 \\ +(-2976\lambda + 80 - 984\lambda^2)k^2 + (1779\lambda - 4536\lambda^3 - 12660 + 13017\lambda^2)k \\ +38400 + 5940\lambda^4 - 75480\lambda^2 - 61020\lambda^3 + 29160\lambda \end{Bmatrix},$$

$$c_3 = \begin{Bmatrix} (-240\lambda - 190)k^3 \\ +(560 + 1542\lambda^2 + 3123\lambda)k^2 + (-1779\lambda + 4536\lambda^3 - 13017\lambda^2 + 12660)k \\ -8910\lambda^4 + 91530\lambda^3 + 113220\lambda^2 - 43740\lambda - 57600 \end{Bmatrix},$$

$$c_4 = \begin{Bmatrix} (-400 - 700\lambda^2 - 1090\lambda)k^2 + (-1512\lambda^3 + 593\lambda + 4339\lambda^2 - 4220)k \\ +38400 + 5940\lambda^4 - 75480\lambda^2 - 61020\lambda^3 + 29160\lambda \end{Bmatrix},$$

$$c_5 = (-9600 + 18870\lambda^2 - 7290\lambda - 1485\lambda^4 + 15255\lambda^3),$$

$$c_6 = 30 \begin{Bmatrix} (-5\beta - 5\beta\lambda + \beta k + 5 + 5\lambda)(\beta k - 3\beta\lambda - 2\beta + 3\lambda + 2)^2 \\ (-7\beta - 7\beta\lambda + \beta k + 7 + 7\lambda) \end{Bmatrix} R^4, \quad (A10)$$

## APPENDIX B: EXPRESSION OF THE CONSTANT COEFFICIENTS PRESENT IN EQUATION (39)

The constant coefficients present in the expression of surfactant concentration in limiting case of $Pe_s \gg 1$, are given by

$$\Gamma_{1,1}^{(Ca)} = \frac{e}{4R^4}\left(\frac{24\lambda\beta - 12\lambda\beta^2 + 5\beta + 7\beta^2 - 12\lambda - 12}{\beta^2}\right),$$

$$\Gamma_{2,0}^{(Ca)} = \frac{5}{42R^4}\left\{\frac{\begin{pmatrix} -63\lambda\beta e^2 + 28\lambda\beta^2 e^2 + 2\beta^2 e^2 \\ +35\lambda e^2 - 37\beta e^2 - 7\beta^2 + 7\beta + 35e^2 \end{pmatrix}}{\beta^2}\right\},$$

$$\Gamma_{2,2}^{(Ca)} = \frac{5e^2}{252R^4}\left(\frac{63\lambda\beta - 28\lambda\beta^2 - 22\beta^2 - 35\lambda + 57\beta - 35}{\beta^2}\right),$$

$$\Gamma_{3,1}^{(Ca)} = \frac{7e}{72R^4}\left(\frac{-13\lambda\beta + 6\lambda\beta^2 - 25\beta^2 + 7\lambda + 18\beta + 7}{\beta^2}\right),$$

$$\Gamma_{4,0}^{(Ca)} = \frac{1}{77R^4}\left(\frac{550\beta e^2 + 91\beta - 550e^2 - 91}{\beta}\right),$$

$$\Gamma_{4,2}^{(Ca)} = \frac{25e^2}{42R^4}\left(\frac{1-\beta}{\beta}\right),$$

(B1)